\documentclass{article}
\pdfoutput=1
\usepackage{jcappub}

\usepackage{aas_macros}
\usepackage{subcaption}
\usepackage{graphicx}%rotating, graphicx,dsfont}
\usepackage{xcolor}
\usepackage{amssymb,amsmath}
\usepackage{hyperref}
\usepackage{multirow}
\usepackage{amsmath,amssymb}
\usepackage{mydefs}

\newcommand{\tj}[6]{ \begin{pmatrix}
   #1 & #2 & #3 \\
   #4 & #5 & #6 
  \end{pmatrix}}

\newcommand{\bea}{\begin{eqnarray}}
\newcommand{\eea}{\end{eqnarray}}
\newcommand{\bdm}{\begin{displaymath}}
\newcommand{\edm}{\end{displaymath}}

\renewcommand{\vec}{\mathbf}

\def\kMpc{\, h \, {\rm Mpc}^{-1}}

\newcommand{\eq}[1]{Eq.~(\ref{#1})}

\newcommand{\fig}[1]{Figure~\ref{#1}}

\newcommand{\vb}[1]{\mathbf{#1}}

\def\fnl{f_{\mathrm{NL}}^{\mathrm{loc}}}

\def\bi#1{\hbox{\boldmath{$#1$}}}
\title{Redshift-weighted constraints on primordial non-Gaussianity from
      the clustering of the eBOSS DR14 quasars in Fourier space}
\author[a,c]{Emanuele Castorina,}
\author[b]{Nick Hand,}
\author[a,b,c]{Uro{\v s} Seljak,}
\author[d]{Florian Beutler,}
\author[e,f]{Chia-Hsun Chuang,}
\author[g]{Cheng Zhao,}
\author[h]{H\'ector Gil-Mar\'in,}
\author[i,j]{Will J. Percival,}
\author[k]{Ashley J. Ross,}
\author[l]{Peter Doohyun Choi,}
\author[m]{Kyle Dawson,}
\author[n]{Axel de la Macorra,}
\author[l]{Graziano Rossi,}
\author[d]{Rossana Ruggeri,}
\author[o]{Donald Schneider,}
\author[p]{Gong-Bo Zhao}

\affiliation[a]{Berkeley Center for Cosmological Physics, Department of Physics, \\ University of California, Berkeley CA 94720}
\affiliation[b]{Department of Astronomy, University of California, Berkeley, CA 94720}
\affiliation[c]{Lawrence Berkeley National Laboratory, 1 Cyclotron Road, Berkeley, CA 93720}
\affiliation[d]{Institute of Cosmology \& Gravitation, University of Portsmouth, Dennis Sciama Building, Portsmouth, PO1 3FX, UK}
\affiliation[e]{Leibniz-Institut fur Astrophysik Potsdam (AIP), An der Sternwarte 16, D-14482 Potsdam, Germany}
\affiliation[f]{Kavli Institute for Particle Astrophysics and Cosmology, Stanford University, 452 Lomita Mall, Stanford, CA 94305, USA}
\affiliation[g]{Laboratoire d'Astrophysique, École Polytechnique Fédérale de Lausanne, 1015 Lausanne, Switzerland}
\affiliation[h]{ICC, University of Barcelona, IEEC-UB, Mart\'i i Franqu\`es, 1, E08028 Barcelona, Spain}
\affiliation[i]{Department of Physics and Astronomy, University of Waterloo, 200 University Ave. W., Waterloo ON N2L 3G1, Canada	}
\affiliation[j]{Perimeter Institute for Theoretical Physics, Waterloo, Ontario N2L 2Y5, Canada}
\affiliation[k]{Department of Astronomy, The Ohio State University, 140 W. 18th Ave., Columbus, OH 43210, USA}
\affiliation[l]{Department of Physics and Astronomy, Sejong University, 209, Neungdong-ro, Gwangjin-gu, Seoul, South Korea	}
\affiliation[m]{Department of Physics and Astronomy, University of Utah, 115 S. 1400 E., Salt Lake City, UT 84112, USA}
\affiliation[n]{Instituto de Física, Universidad Nacional Autónoma de México, Apdo. Postal 20-364, México}
\affiliation[o]{Institute for Gravitation and the Cosmos, Pennsylvania State University, University Park, PA 16802, USA}
\affiliation[p]{National Astronomical Observatories of China, Chinese Academy of Sciences, 20A Datun Road, Chaoyang District, Beijing 100012, China}

\abstract{
We present constraints on local primordial non-Gaussianity (PNG), parametrized through $\fnl$, using
the Sloan Digital Sky Survey IV extended Baryon Oscillation Spectroscopic
Survey Data Release 14 quasar sample. We measure and analyze
the anisotropic clustering of the quasars in Fourier space, testing for
the scale-dependent bias introduced by primordial non-Gaussianity on
large scales. We derive and employ a power spectrum estimator using
optimal weights that account for the redshift evolution of the PNG signal.
We find constraints of $-51<\fnl<21$ at 95\% confidence level. These are amont the tightest constraints from Large Scale Structure (LSS) data.
Our redshift weighting improves the error bar by 15\% in comparison to the unweighted case. If quasars have lower response to PNG, the constraint degrades to $-81<\fnl<26$, with a 40\% improvement over the standard approach.
We forecast that the full eBOSS dataset could reach $\sigma_{\fnl}\simeq 5\text{-}8$ using optimal methods and full range of scales.
}

\begin{document}
\maketitle

\section{Introduction}

Measurements of the statistical properties of the late-time large-scale
structure (LSS) of the Universe can provide insights into the
physics that generated the primordial density fluctuations.
In particular, they offer the possibility to distinguish between
different models of cosmic inflation by measuring primordial non-Gaussianity (PNG),
the deviation from Gaussian random field initial conditions.
In this work, we focus on the local type of PNG, through
the one parameter $\fnl$. Single-field inflationary models predict
an amplitude of $\fnl$ that is unmeasurably small, and a detection
of $|\fnl| \gtrsim 1$ would robustly rule out this class of
inflationary models \cite{Maldacena:2003,Creminelli:2004}.

The current state-of-the-art constraint on PNG comes not from LSS data but
from measurements of the bispectrum of the cosmic microwave
background (CMB) by the \textit{Planck} satellite, which has reported
$\fnl = 0.8 \pm 5.0$ \cite{Planck:2016}. Unfortunately, the
improvement in precision from CMB measurements is not expected to reach
the level required to distinguish between
inflationary models ($\sigma(\fnl) \sim 1$) due to cosmic variance limitations
\cite{Baumann:2009,Abazajian:2016}. However, forecasts for future
LSS surveys, e.g., \cite{Dore:2014,Yamauchi:2014,Ferramacho:2014,
Ferraro:2015,Raccanelli:2015,Camera:2015,Alonso:2015,Tucci:2016,
dePutter:2017,Karagiannis}, indicate a strong potential for PNG constraints.
The sensitivity to PNG originates from the
distinctive scale-dependent bias signature that is imprinted on the
clustering of biased tracers (e.g., galaxies or quasars) by local
primordial non-Gaussianity
\cite{Dalal:2008,Matarrese:2008,Slosar:2008,Desjacques:2010}
(see e.g. \cite{Alvarez:2014} for a review).
The effect is proportional to the bias of the tracers themselves
and scales as $\fnl k^{-2}$; thus, it is most prominent on the largest scales probed by a survey.
Further gains can be made by surveys that observe multiple tracers,
which are able to effectively remove uncertainties from sample variance in their
measurements \cite{Seljak:2009,McDonald:2009,Hamaus:2011,Castorina_zerobias}.

The current best constraints from the analysis of large-scale structure
data are comparable to those found by the \textit{WMAP} CMB experiment
\cite{Slosar:2008,Ross:2013,Giannantonio:2014b,Leistedt:2014,Ho:2015,Bennett:2013}. The first such analysis by
\cite{Slosar:2008} combined a number of different tracers from early SDSS
releases to find $\fnl = 28^{+23}_{-24}$ (68\% CL).
This analysis also demonstrated the constraining power of quasars, finding
$\fnl = 8^{+26}_{-37}$ at 68\% CL using only the SDSS photometric quasar
sample. For recent constraints using QSOs see \cite{Giannantonio:2014b,Leistedt:2014,Karagiannis14}.
As quasars are highly biased and
probe large volumes, they are ideal for measuring the PNG signal on large
scales. On the other hand large scales are the most contaminated by systematic effects
\cite{Ross:2012,Ross:2013,Pullen:2013,Leistedt:2013,Leistedt:2014b,Ho:2015,Kalus19}.
Systematics control has spurred work on the use of cross-correlations
in LSS PNG analyses, e.g., \cite{Giannantonio:2014,Schmittfull:2017}.

Data sets that probe large volumes offer the best chance to detect
non-Gaussian biasing features on large scales, but they also complicate
data analysis. For samples that span a wide redshift range, traditional
analysis methods, such as using multiple, smaller redshift bins,
become non-optimal. A proper treatment of the redshift evolution of
the tracer bias and PNG signal is therefore necessary to fully exploit the
constraining power of a data set. Recent work has
focused on using redshift weights to optimize LSS surveys for baryon acoustic
oscillation (BAO) and redshift-space distortion (RSD) analyses
\cite{Zhu:2015,Zhu:2016,Ruggeri:2017}. The methods presented in the aforementioned works have been recently applied to the first data release and cosmological analyses of the extended
Baryon Oscillation Spectroscopic Survey (eBOSS; \cite{Dawson:2016}) survey \citep{DR14Ruggeri18RSD,DR14Zhu18BAO,Zhao:2017,DR14Zarrouk18RSD,DR14Wang18BAO}. The idea of redshift weighting scheme was also
extended in \cite{Mueller:2017} to optimize for PNG constraints.
The purpose of this work is to  present and clarify the methodology to perform an optimal, in a statistical sense, signal weighted measurement of PNG using galaxy surveys data. 
% As an application of our method, we will present constraints on PNG using the distribution of QSOs in the DR14 of eBOSS data. 
% However, we think a few key points make those analysis not optimal in a statistical sense. In \cite{Zhu:2015,Zhu:2016} the redshift evolution of the underlying cosmological signal was not included in the optimal weights. In the Fourier space methods of \citep{Ruggeri:2017,Mueller:2017} the optimal weights have been derived for pairs of galaxies in configuration space, which cannot therefore be applied the power spectrum multipoles. 

Our first goal is to derive a redshift-weighted optimal quadratic
estimator for the two-point statistics that yields optimal
constraints for $\fnl$. Our method is general and can be applied to any other parameter, using measurements in configuration space or Fourier space in a spectroscopic or photometric catalog.
We will also show that optimal redshift weights to a good approximation change the effective redshift of a survey, in a way that is completely analogous to the standard FKP weights \citep{Feldman:1994}.
As an application of our method we use the Sloan Digital Sky Survey (SDSS) IV eBOSS
Data Release 14 quasar sample (DR14Q) \citep{eBOSS-DR14} to derive constraints on $\fnl$. This data set
includes 148,659 quasars and spans a redshift range of $0.8 \le z \le 2.2$.

This paper is organized as follows. We present our new
optimal estimator, which correctly accounts for redshift evolution of
the signal, in \S\ref{sec:theory}. In \S\ref{sec:data} we describe
the eBOSS quasar sample used in this work. 
 \S\ref{sec:methods} outlines our analysis methods, including how we estimate the power spectrum
multipoles of the data and the theoretical model used to estimate parameters.
In \S\ref{sec:Fisher} we study the Fisher matrix of eBOSS data to try to quantify a-priori the improvement yielded by the optimal analysis.
We present our constraints on $\fnl$ in \S\ref{sec:constraints} and discuss and conclude
in \S\ref{sec:conclusions}.

\section{Primordial Non-Gaussianities in the Large Scale Structure}\label{sec:theory}

\subsection{Local PNG}

In this work we focus on the local type of primordial non-Gaussianity,
where the primordial potential, $\Phi_p(\mathbf{x})$, is the sum of a random Gaussian field, $\phi$, and
its square,

\begin{equation}
  \Phi_p(\mathbf{x}) = \phi(\mathbf{x})  + \fnl \left (\phi(\mathbf{x}) ^2 - \la \phi^2 \ra \right),
\end{equation}
with $\fnl$ parametrizing the amount of PNG.
The relation between $\Phi_p$ and the matter overdensity $\delta_m$
is easiest to express in Fourier space, where it is given by
$\delta_m(k,z) = \alpha(k,z) \Phi_p(k)$, with

\begin{equation}\label{eq:alpha}
  \alpha(k,z) = \frac{2 c^2 k^2 T(k) D(z)}{3 \Omega_m H_0^2}
\end{equation}
where $T(k)$ is the transfer function, $c$ is the speed of light, $D(z)$ is
the linear growth factor normalized to $(1+z)^{-1}$ in the matter-dominated
era, $\Omega_m$ is the matter density parameter
at $z=0$, and $H_0$ is the present-day
Hubble parameter. We also define a related quantity $\atilde$, which will be
useful in the discussion to follow:

\begin{equation}\label{eq:atilde}
  \atilde(k,z) \equiv \frac{2\deltac}{\alpha(k,z)} = \frac{3 \Omega_m H_0^2 \deltac}{c^2k^2T(k)D(z)},
\end{equation}
with $\deltac = 1.686$, the critical density in the spherical collapse
model in a Einstein-deSitter Universe.

As shown in \cite{Dalal:2008,Slosar:2008,Desjacques:2010}, local PNG as
parametrized by $\fnl$ introduces a scale-dependent bias, $\deltab(k,z)$,
given by

\begin{equation}
\label{eq:def_bias}
  \deltab(k,z) \equiv  b_\phi \fnl \atilde(k,z) = (b-p)\fnl \atilde(k,z),
\end{equation}
where $b_\phi$ is the response of the halo or galaxy field to the presence of PNG. In the last equality we assumed $b_\phi= (b-p)$ where $b$ is the bias of the sample. This is an approximation to the exact $\fnl$ response of discrete tracers, but measurements in N-body simulations have shown it is good enough to estimate the amplitude of scale-dependent bias \citep{Biagetti16}.
The parameter $p$  takes a value of 1 for a halo mass selected sample and 1.6 for samples
dominated by recent mergers \cite{Slosar:2008,Reid}, as could be the case for QSOs for instance. For the purpose of deriving an optimal estimator we will not fix $p$ for now.
At the linear order, and after adding redshift-space distortions
(e.g., \cite{Kaiser:1987}), we find the quasar overdensity is related to the
matter overdensity in the presence of PNG as follows

\begin{equation} \label{eq:deltaQ-PNG}
\delta_\qso = [b + f \mu^2 + \deltab] \delta_m \equiv [\btilde + \deltab]\delta_m
\end{equation}
where $f = d\ln D / d\ln a$ is the logarithmic growth rate, $\mu$ is the cosine of the angle between the Fourier modes and the line of sight, and we
have defined the convenient quantity $\btilde = b + f \mu^2$, which accounts
for both Gaussian biasing and linear redshift-space distortions.

\subsection{Optimal estimators in LSS}\label{sec:optimal-lss}

Our goal is to derive an estimator for the two-point clustering of
a data set that yields the tightest constraint on $\fnl$. We begin by
describing the data, positions on the sky and redshifts of set of objects, in terms of the pixelized overdensity
$\delta_\qso(\vri)$, where $\vri$ gives the pixel position. We will also need the
mean density at a given pixel position, denoted as $\nbar$.
Optimal analysis invariably requires inverse noise weighting of the data.
For example, if $\nbar=0$ then no data have been observed at
that pixel and it should not be used for data analysis, suggesting that the
noise should be infinite. An additional source of uncertainty is
sample variance, which is caused by the finite number of
measureable modes and is present even in absence of noise.

When considering Gaussian statistics, the optimal inverse noise
weighting of a data set has a well-defined solution, known as
the optimal quadratic estimator \cite{Tegmark:1997c,Bond:2000}, which
weights the data inversely by the covariance matrix. If
we collect our overdensity pixels into a vector $\vx$,
with $x_i = \delta_\qso(\vri)$, then its signal covariance matrix is $S_{ij}$,
and the total covariance matrix reads

\begin{equation}
C_{ij} = \la x_i x_j \ra = N_{ij} + S_{ij} = 
       = [V \nbar]^{-1} \kronecker + S_{ij},
\label{eq:Cij}
\end{equation}
where $\kronecker$ is the Kronecker delta, $V$ is the pixel volume, and
we have assumed Poisson statistics for the noise term $N_{ij}$.

The optimal quadratic estimator (OQE) for a parameter $\theta$
is then \cite{Tegmark:1997b,Tegmark:1997,Tegmark:1998b,Abramo:2016}

\begin{equation}
\qhat_\theta = \frac{1}{2} \vx^t C^{-1} C_{,\theta} C^{-1} \vx - \Delta q_\theta,
\label{eq:optimal-quadratic-estimator}
\end{equation}
where $C_{,\theta}$ denotes the derivative of $C$ with respect to $\theta$,
and $\Delta q_\theta$ subtracts a possible bias of the estimator.
If the response of the covariance matrix is constant in $\theta$, then the OQE is also the maximum likelihood solution.

The most difficult task is to compute $C^{-1}\vx$, and a diagonal form for the configuration space covariance matrix $C$ is often employed to evaluate $C^{-1}$.
Suppose indeed we want to determine, using \eq{eq:optimal-quadratic-estimator}, the power spectrum around some $k$, where we expect
 the power to be close to a fiducial power spectrum $\Pfid$.
If we assume that the power spectrum is locally
flat (white noise) within the band powers then its Fourier transfer would be a zero lag correlation
function determined by the amplitude of the power spectrum. This gives rise to
a diagonal inverse of the covariance matrix in configuration space,
\begin{equation}
C_{ij}^{-1} = (\Pfid (k) + \bar{n}^{-1})^{-1} V \kronecker\,.
\label{eq:C_fkp}
\end{equation}
The fiducial power spectrum should in principle be varied with $k$,
but over the range of scales one is usually interested in it is a relatively slowly varying function. For PNG, we are concerned with the power
on the largest scales, and we can assume a constant fiducial value $\Pfid \sim 3 \times 10^4 \hMpcc$ for all wavenumbers $k$.

We also need to evaluate the derivative $C_{,\theta}$, where
$\theta$ is the parameter we wish to determine, in our case the averaged value of the power spectrum. Suppose we focus first
on a single mode $k$ with a volume $\mathrm{d}\vk = (2\pi)^3/V$. The
Fourier transform of the power spectrum is the correlation function,
which for this single mode gives
$S_{ij} = V^{-1}P(k) \exp[i \vk (\vri - \vrj)]$.
Its derivative with respect to $P(\bi{k})$ gives

\begin{equation}
\frac{dC_{ij}}{dP(k)} = V^{-1} e^{i \vk (\vri - \vrj)},
\end{equation}
and the estimator of \eq{eq:optimal-quadratic-estimator}
for the power spectrum becomes

\begin{equation}\label{eq:fkp-estimator}
\hat{P}(k) = A \left\vert \sum_j e^{i \vk \vrj} \wfkp \right\vert^2,
\end{equation}
where we have replaced the sum over pixels with the sum over discrete objects,
such that $\delta_\qso\bar{n}V=N_\qso$, where $N_\qso$
is the number of objects in the pixel (if the pixels are small enough
this can be viewed just as the sum over object positions $\vri$).
The weights $\wfkp$ take the well-known form as first derived in
\cite{Feldman:1994}, $\wfkp=(1+\bar{n}\Pfid)^{-1}$.
We see that the operation in \eq{eq:fkp-estimator} is a
Fourier transform, which can be computed rapidly using fast Fourier transforms
(FFTs). The normalization $A$ can be determined
by performing the same operation on an unclustered catalog of synthetic
objects, including FKP weights, and normalized to the total number of
observed objects
\cite{Feldman:1994,Yamamoto:2006,Bianchi:2015,Scoccimarro:2015,Hand:2017}.

\subsection{Optimal estimator for $\fnl$}\label{sec:fnl-weights}

Now we consider instead the weighting scheme that yields an optimal
constraint on $\fnl$. We explicitly account for redshift evolution
by considering overdensity pixels as a function of
time, or redshift, $\vr = \vr(z)$. We begin by computing the signal covariance in the
presence of PNG from \eq{eq:deltaQ-PNG},

\begin{align}
S_{12} &= \la \delta_\qso(\vrone(z_1)) \delta_\qso(\vrtwo(z_2)) \ra \\
       &=
\la
  \left[
      (\btilde_1 + \deltab_1) \delta_m(\vrone)
  \right]
  \left[
      (\btilde_2 + \deltab_2) \delta_m(\vrtwo)
  \right]
\ra,
\end{align}
where we have defined  $\btilde_1 = \btilde(z_1)$,
$\vrone = \vr(z_1)$, $\deltab_1 = \deltab(z_1)$, and similar quantities
at $z_2$. Evaluating the derivative of this expression at $\fnl=0$ yields

\begin{equation}\label{eq:S12-start}
\frac{d S_{12}}{d\fnl}\Biggr|_{\fnl=0} =
    \btilde_1 (b_2 - p) \atilde_2
        \la \delta_m(\vrone) \delta_m(\vrtwo) \ra
    + \ \mathrm{1 \leftrightarrow 2 }\;,
\end{equation}
where the second term is symmetric and can be computed via an exchange of
indices. We can use the definition of the the power spectrum
to express this equation as

\begin{equation}\label{eq:S12-final}
\frac{d S_{12}}{d\fnl}\Biggr|_{\fnl=0} =
  (b_1 - p)b_2 \int \frac{d\vk}{2\pi^3}
                \atilde_1(k) (1 + \beta_2\mu_{r_2}^2) P_m(k, z_1, z_2) e^{i \vk (\vrone-\vrtwo)}
                + \ \mathrm{1 \leftrightarrow 2 },
\end{equation}
where $P_m(k)$ the matter power spectrum, $\beta = f/b$ is the standard
RSD parameter, and $\mu_{r_2}$ is the line-of-sight angle associated with
position $\vrtwo$.

It is useful to factor out some of the time dependencies in
equation \eq{eq:S12-final} using

\begin{align}
P_m(k, z_1, z_2) & = P_m(k,z_0)(k) D(z_1) D(z_2)/D(z_0)^2, \\
\atilde(k, z) & = \frac{\atilde_0(k)}{D(z)}\;,
\end{align}
with $z_0$ some reference time, we adopted $z=0$. \footnote{Multiplicative constants independent of wavenumber $k$ and redshift, like $D(z_0)$, can be safely neglected as the final result is always properly normalized.}
With these definitions, we can
express the optimal estimator in \eq{eq:optimal-quadratic-estimator}
as a function of $\vrone$ and $\vrtwo$ as

\begin{align}
\qhat_{\fnl}(\vrone, \vrtwo) &= \frac{1}{2} C^{-1} \vx^t \frac{d S_{12}}{d\fnl}\Biggr|_{\fnl=0}C^{-1} \vx - \Delta q_{\fnl} \nn \\
    &= \frac{1}{2}\frac{\delta_\qso(\vrone)}{C}
    \Biggr[\int \frac{d\vk}{2\pi^3} e^{i\vk(\vrone-\vrtwo)}
    P_{m,0}(k) \atilde_0(k) D(z_2) (1 + \beta_2 \mu_{\vrtwo}^2)(b_1-p)b_2 \nn \\
    &+ \ \mathrm{1 \leftrightarrow 2 } \Biggr]\frac{\delta_\qso(\vrtwo)}{C}
    - \Delta q_{\fnl}.
\end{align}

\begin{figure}[tb]
\centering
\includegraphics[width=0.5\textwidth]{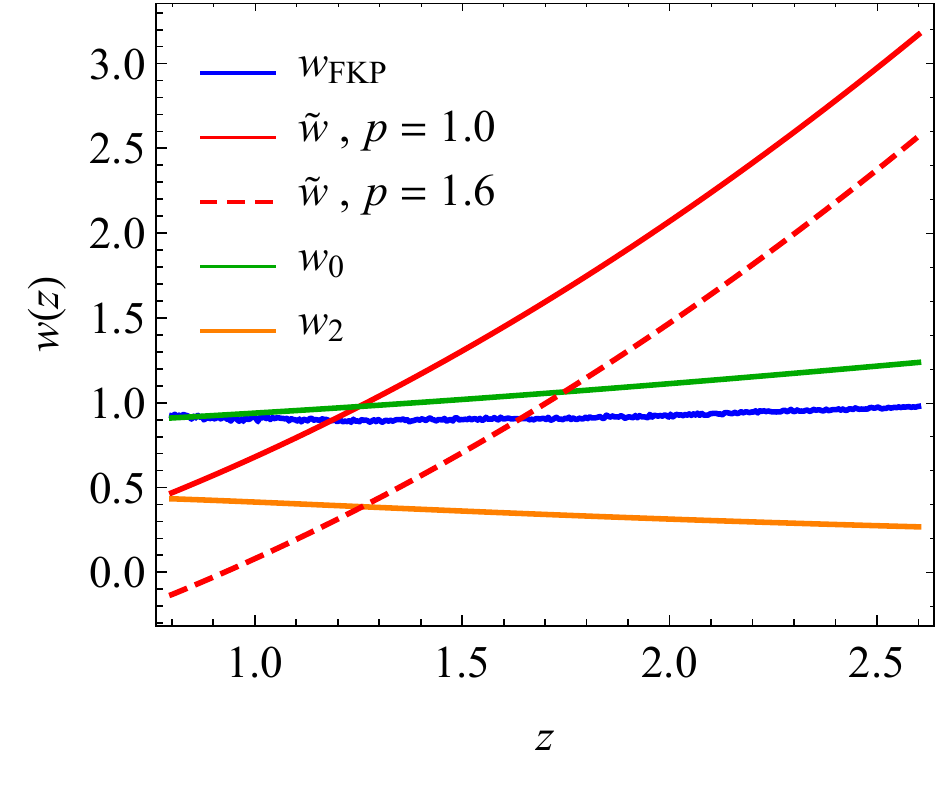}
\caption{Redshift weights for the eBOSS DR14 QSO catalog. Since $n(z)P_0 \ll 1$, the  standard FKP weights, in blue, show small redshift evolution. The optimal weights to estimate PNG are shown in red, green and orange. The weights depend on the QSOs response to $\fnl$, as one can see from the difference between the solid red line, for $p=1$, and the dashed one, for $p=1.6$.}
\label{fig:wz}
\end{figure}

And now, summing over $\vrone$ and $\vrtwo$, we obtain the estimator

\begin{align}\label{eq:qhat-final}
\qhat_{\fnl} &= \frac{1}{2} \int \frac{d\vk}{(2\pi)^3} P_{m,0}(k)\atilde_0(k) \nn \\
      & \Biggr\{ \left[ \int d\vrone \ e^{i\vk\vrone} \frac{\delta_\qso(\vrone)}{C} (b_1-p)\right]
      \left[ \int d\vrtwo \ e^{-i\vk\vrtwo} \frac{\delta_\qso(\vrtwo)}{C} b_2 D(z_2) (1 + \beta_2 \mu_{\vrtwo}^2))\right] \nn \\
      &+ \ \mathrm{1 \leftrightarrow 2 } \Biggr\} - \Delta q_{\fnl}.
\end{align}
Note that in these equations the inverse noise weight factors of $C^{-1}$ are
identical to those discussed in Section~\ref{sec:optimal-lss}, with the
FKP weight being the near-optimal scheme. We can further decompose in Legendre polynomials the angular part  of the Kaiser factor,
\begin{align}
(1+ \beta \mu^2) = \left(1+\frac{\beta}{3}\right)\mathcal{L}_0(\mu) + \frac{2}{3}\beta\mathcal{L}_2(\mu)
\end{align}
then define, for a generic weight $w(z)$, the weighted density multipoles
\begin{equation}
\delta_\ell^w(\mathbf{k}) = \int \mathrm{d}^3r\,e^{i\mathbf{k}\cdot \mathbf{r}} w_{\rm FKP}(z) w(z) \delta (\mathbf{r}) \mathcal{L}_\ell(\hat{k}\cdot\hat{r})
\end{equation}
to finally obtain 
\begin{align}
\label{eq:estimator}
\qhat_{\fnl} &=  \int \frac{\mathrm{d}k\,k^2}{2\pi^2} P_{m,0}(k)\atilde_0(k) \int \frac{d \Omega_k}{4\pi}[\delta_{0}^{\tilde{w}}(-\mathbf{k})\sum_{\ell=0,2} \delta_{\ell}^{w_{\ell}}(\mathbf{k})] - \Delta q_{\fnl}
\end{align}
where \begin{align}
\tilde{w}(z) = b(z)-p\quad,\quad w_0(z) =D(z)(b(z)+f(z)/3)\quad,\quad w_2(z) =2/3 D(z)f(z)\;.
\end{align}
The above \eq{eq:estimator} and the associated set of weights defines the optimal signal weighting of the power spectrum, and they represent one of the main results of this work.
One can immediately recognize that, if we neglect the optimal weights, the structure of angular integral over the wavenumber $\vk$ is the same of standard estimators of the monopole and quadrupole of the power spectrum \citep{Yamamoto:2006,Bianchi:2015,Scoccimarro:2015,Hand:2017,BeutlerCastorina}.
This is expected, as the optimal redshift weights are no different than FKP weights in this respect.
Within our model, the hexadecapole, $\ell=4$, does not carry any information about PNG and therefore does not appear in the optimal estimator. %In practice though, if one needs to marginalize over several nuisance parameters higher order multipoles could be included.
We notice that optimal weights for PNG up-weight high redshift galaxies, which are highly biased and have therefore larger $\fnl$ response. The OQE also exploits the fact that the primordial potential does not evolve in redshift, whereas the Gaussian part of the signal does. This can be seen by comparing the different dependence of $\tilde{w}$ and $w_\ell$ on the linear growth function $D(z)$. Figure~\ref{fig:wz} shows the redshift evolution of the unormalized weights in the eBOSS DR14 QSOs catalog, described in more details in Section \ref{sec:data}. For the DR14 sample, $w_{\rm FKP}(z)$, blue line, and $w_\ell(z)$, green and orange lines,  slowly vary across the survey, the former since $n(z)P_0 \ll 1$ whereas the latter because $b_{\rm qso}(z)D(z)\simeq \rm{const.}$
One the other hand $\tilde{w}(z)$ grows quite rapidly with resdhift, up-weighting galaxies with a larger response to $\fnl$. 

Finally, the estimator in \eq{eq:estimator} in principle needs to be made
unbiased by subtracting out the signal in the absence of
any $\fnl$ via the $\Delta q_{\fnl}$ term. 
As we will see in the next section, optimal weighting boils down to a redefinition of the mean redshift of the survey, allowing us to use standard tools to constrain $\fnl$ or other parameters, and obtain statistically unbiased results.
It is also straightforward to generalize the estimator defined above to the case of cross-correlation between different tracers.

We conclude this section with a few remarks. 
Our estimator differs from the one in \citep{Mueller:2017}, in which the authors defined optimal weights for pairs of galaxies, under the assumption that a single redshift can be associated to each pair. The drawback of this approach is that there is no straightforward implementation in Fourier Space, as there is no Fourier space analog of a pair of objects.
This led \citep{Mueller:2017} to define weights for individual galaxies as a square root of pair weights. However, weights can be negative and one is forced to take an absolute value before the square root. 
We'd like to stress that configuration and Fourier space carry the same amount of information, and as such the same weighting scheme should apply to both. 
Our method naturally addresses this issue, as the optimal weights are defined for each individual object. The same arguments hold for the weights derived in \citep{Ruggeri:2017} for redshift-space distortions parameters and for BAO in \cite{Zhu:2016}.

\section{Data}\label{sec:data}

In this section, we describe the eBOSS DR14Q sample and the synthetic
mock catalogs used in our analysis.

\subsection{eBOSS DR14Q sample}

The extended Baryon Oscillation Spectroscopic Survey \cite{Dawson:2016}
is part of the SDSS-IV experiment \cite{Blanton:2017}.  The eBOSS cosmology
program relies on the same optical spectrographs \cite{Smee:2013} as the
SDSS-III BOSS survey, installed on the 2.5 meter Sloan Foundation
Telescope \cite{Gunn:2006} at the Apache Point Observatory in New Mexico.
In addition to observing luminous red galaxies and emission line
galaxies, eBOSS will observe and measure
redshifts for $\sim$500,000 quasars across a volume of the Universe unexamined
by previous spectroscopic surveys. First eBOSS cosmology results for the DR14Q
sample were recently presented in \cite{Ata:2017}, which reported
the first BAO distance measurement in the range $1 < z < 2$.
The clustering properties of the eBOSS quasars have also
been previously examined in \cite{Laurent:2017,Rodriguez-Torres:2017},
although these works do not make use of the full DR14Q sample.
Recent work in \citep{DR14Ruggeri18RSD,DR14Zhu18BAO,Zhao:2017,DR14Zarrouk18RSD,DR14Wang18BAO,DR14Hou18RSD,DR14Gil-Marin18RSD} has presented several application of eBOSS DR14 data to measurements of BAO and redshift space distortions parameters.

The imaging data, target selection, and catalog construction
methods for the DR14Q sample used in this work are discussed in detail in
\cite{Paris:2017,eBOSS-DR14}. Targets are selected from the catalogs of
the SDSS I/II surveys \cite{York:2000}, released as part of SDSS DR7
\cite{Abazajian:2009}, and the SDSS-III survey \cite{Eisenstein:2011,Dawson:2013},
released as part of SDSS DR8 \cite{Aihara:2011}. The eBOSS also
makes use of several bands of the Wide Field Infrared Survey Explorer
(WISE; \cite{Wright:2010}), as described in \cite{Myers:2015}.
The target selection criteria for the DR14Q sample is presented in detail in
\cite{Myers:2007,RossNP2012}.

Accurate redshift estimation is crucial for achieving the cosmology goals
of eBOSS, which is particularly challenging for quasar spectra
\cite{Shen:2016}. As described in \cite{Paris:2017}, the DR14Q sample contains
three automated redshift estimates per object. In this work, we use the
so-called ``fiducial'' redshift $z_\mathrm{fid}$, which can be any of the three
redshift estimates, depending on which one yields the lowest catastrophic
failure rate (see \cite{Paris:2017} for further details).

\begin{figure}
  \centering
  \includegraphics[width=0.8\textwidth]{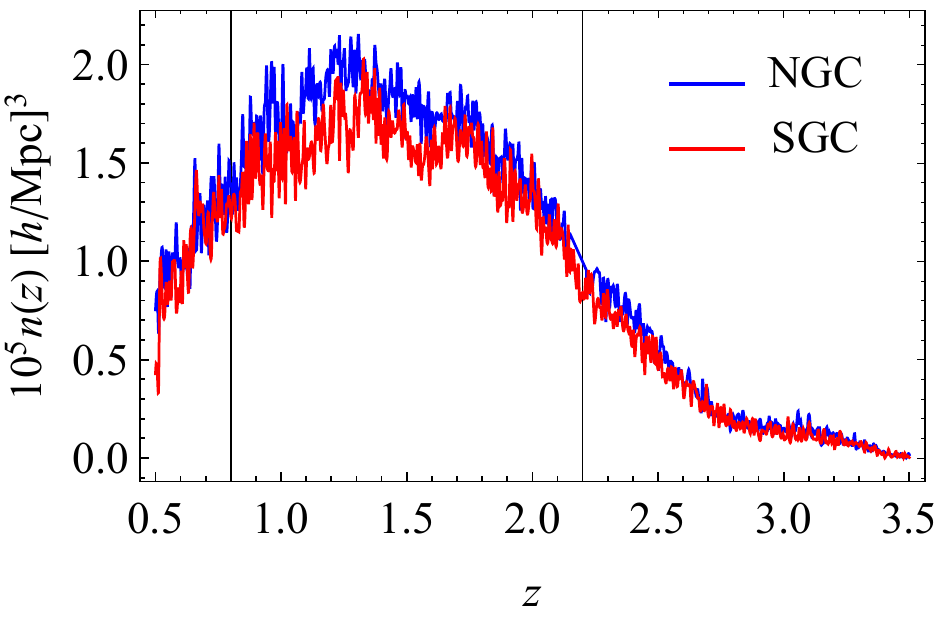}
  \caption{The mean density of quasars in the DR14Q sample as a function of
  redshift for the NGC (blue) and SGC (red) regions of the sky. The
  differences between the two regions are due to known discrepancies
  with the targeting efficiency. The two vertical lines bracket the DR14 QSOs in $0.8<z<2.2$.}
  \label{fig:nbar}
\end{figure}

The DR14Q sample contains 148,659 objects with spectroscopic redshifts
in the range $0.8 \le z \le 2.2$. The observed objects are
distributed in two separate angular regions in the North Galactic Cap (NGC)
and South Galactic Cap (SGC). The effective areas of these regions are
1214.6 $\mathrm{deg}^2$ and 899.3 $\mathrm{deg}^2$, respectively.
We show the observed number density as a function of redshift for the
NGC and SGC regions in Figure~\ref{fig:nbar}. There are slight discrepancies
in $n(z)$ between the two regions due to differences in targeting
efficiency.

\subsection{Completeness weights}\label{sec:comp-weights}

Objects in the DR14Q sample are assigned weights to account for the
incompleteness of the target selection process and other systematic effects
that could potentially bias our clustering measurements. There are two
main types of weights that we will discuss in this section,
spectroscopic completion weights $\wspec$ and systematic
imaging-based weights $\wsys$.
The former accounts for the fact that a small percentage of
targets do not receive a redshift while the latter set of weights corrects
for systematics arising from photometric inhomogeneities in the
targeting selection. When combining these two sets of weights, we take
the total completeness weight as

\begin{equation}\label{eq:wc}
    \wc = \wsys \cdot \wspec.
\end{equation}

\subsubsection{Spectroscopic weights}\label{sec:spec-weights}

The first main cause of spectroscopic incompleteness in the DR14Q
sample is \textit{fiber collisions}. Fiber collisions
result when a pair of quasars are separated by less than the
62\arcsec angular size of the SDSS spectrograph fiber, which prevents
one of the objects from being observed. Missed observations are
partially alleviated by the eBOSS tiling pattern, which naturally has
overlapping tiles in regions with a density of targets on the sky, and
thus, allows redshifts to be measured for objects separated by less than
the 62\arcsec collision scale. Ultimately, 4\% and 3\% of the eBOSS
quasar targets are fiber-collided objects that do not receive
a spectroscopic observation in the NGC and SGC regions, respectively.

We account for the missing objects due to fiber collisions by up-weighting
the nearest neighbor with a valid redshift and spectroscopic class. This
procedure follows previous clustering analyses, e.g., as in
BOSS \cite{Anderson:2014, Reid:2014}. In practice, this is not a perfect
correction, as a fraction of fiber collision pairs are mere projections
and are not associated with the same dark matter halo. However,
the nearest neighbor weighting scheme does preserve the large-scale
bias of the clustering sample. As we are concerned only with the
PNG signal on large scales, we leave exploration of most advanced fiber
collision correction schemes, e.g., \cite{Hahn:2017}, for future work.
We denote the weight used to correct for fiber collisions as a
\textit{close pair} weight, $\wcp$. By
default, its value is unity for all objects that are not involved in
a fiber collision, and for the case of fiber collisions, it is
equal to an integer with value greater than unity.

The second main cause of spectroscopic incompleteness is
\textit{redshift failures}, which refers to the subset of quasars
that do not receive a valid automated redshift
and are not visually inspected. The distribution of these objects is not
uniform within the focal plane due to variations in detector efficiency.
In past BOSS releases \cite{Reid:2016}, redshift failures were an
almost negligible fraction of the total, less than 1\%. However,
redshift determination for a quasar at $z \sim 1.5$ is more difficult
than for an LRG at $z \sim 0.5$, and the DR14Q sample has a redshift
failure rate of 3.4\% and 3.6\% in the NGC and SGC, respectively.
With this increased rate, a more complex scheme than was used in
previous BOSS analyses, is required to adequately correct for the
effect. 
%This more complicated correction procedure was not used
%in the first eBOSS BAO analysis of \cite{Ata:2017}, but was implemented for
%the RSD analyses \cite{GilMarin:2017-inpre}. 
Here, we use a
focal plane weight $\wfoc$ defined as

\begin{equation}
  \wfoc = \left [ 1 - P_\mathrm{rf}(x_\mathrm{foc}, y_\mathrm{foc})\right]^{-1},
\end{equation}
where $P_\mathrm{rf}$ defines the probability of obtaining a redshift failure
as a function of position in the focal plane. With this weight, quasars
with measured redshifts that are observed in positions on the focal plane
where $P_\mathrm{rf}$ is greater than zero will be up-weighted to account
for the fact that, on average, targeted quasars are missing from the sample
due to redshift failures. We refer the reader to \cite{DR14Gil-Marin18RSD}
for further details on the redshift failure weights.

Finally, we assign the total spectroscopic completeness weight as the
product of the fiber collision and redshift failure weights,
$\wspec = \wcp \cdot \wfoc$.

\subsubsection{Imaging weights}

Each quasar in the DR14Q sample is also assigned a weight to mitigate
photometric systematics, using the prescription studied in
\cite{Laurent:2017}. The weights, denoted here as $\wsys$,
account for inhomogeneities in the quasar targeting selection related
to the Galactic extinction and depth of the targeting image data.
The weights used in this work have been utilized in previous eBOSS
cosmology analyses \cite{Ata:2017,DR14Gil-Marin18RSD}. They are described
in detail in section~3.4 of \cite{Ata:2017}, and we refer the reader to
that work for further details.

\subsection{Synthetic DR14Q catalogs}\label{sec:ezmocks}

We make use of a set of mock catalogs to estimate the covariance matrix of the eBOSS power spectrum measurements. The mocks are based on the Extended
Zel'dovich (EZ) approximate $N$-body simulation scheme \cite{Chuang:2015}.
Throughout this work, we refer to this set of synthetic catalogs as EZ mocks.
In total, we utilize 1000 independent realizations for each Galactic cap
region. We
also use the mocks to verify and test our analysis and parameter
estimation pipelines.

The set of EZ mocks are generated following the methodology outlined
in \cite{Chuang:2015}, matching both the angular footprint and redshift
selection function of the DR14Q sample. Briefly, the EZ mock scheme
relies on the Zel'dovich approximation to generate a density field,
and implements nonlinear and halo biasing effects through the use of
free parameters. These free parameters are tuned to produce two-point
and three-point clustering of a desired data set. The method allows
for the fast generation of a large number of mock catalogs without the
computational cost of full $N$-body simulations, and it has been used
extensively in previous BOSS cosmology analyses, e.g.,
\cite{Kitaura:2016, Alam:2017}.

The EZ mock catalogs account for the redshift evolution of the eBOSS quasars
by constructing a light-cone out of 7 redshift shells, generated
from periodic boxes of side length $L = 5000 \hMpc$ at different redshifts.
The free parameters of each box are calibrated independently, and the
boxes are combined using the \texttt{make\_survey} software
\cite{Carlson:2010}.
The background density field of the light cone mocks is continuous, as
each of the boxes shares the same initial Gaussian density field.
The NGC and SGC data sets are treated independently when deriving the best-fit
internal EZ mock parameters.
The cosmology of the EZ mocks is a flat, $\Lambda\mathrm{CDM}$ model with
$\Omega_m = 0.307115$, $\Omega_b = 0.048206$, $h=0.6777$, $\sigma_8 = 0.8255$,
and $n_s = 0.9611$. 

Finally, we mirror the effects of fiber collisions and redshift failures
(as discussed in section~\ref{sec:spec-weights}) and each object in an
EZ mock catalog also has associated values for $\wfoc$ and $\wcp$.
Fiber collisions are implemented by applying the tiling pattern to the
mock data, and removing pairs that fall within the collision scale that
are not in overlapping tiles. Redshift failures are applied by
statistically removing objects based on the position of the object in
the focal plane, using the probability of a redshift failure
$P_\mathrm{rf}(x_\mathrm{foc}, y_\mathrm{foc})$.

% \begin{figure}[!tb]
%   \centering
%   \includegraphics{figures/ezmock_comparison.pdf}
%   \caption{The measured power spectrum multipoles for the DR14Q sample,
%   as compared to those measured from the mean of the 1000 EZ mock catalogs.
%   We show the comparison separately for the NGC (top) and SGC (bottom) data sets.
%   Errors on the data measurements are computed from the variance of the
%   1000 EZ mock measurements. We do not show error bars on the mean of the
%   EZ mock multipoles (black), as they are small compared to that of the data.}
%   \label{fig:ezmock-comparison}
% \end{figure}

\section{Analysis Methods}\label{sec:methods}
Throughout our analysis, we assume a flat $\Lambda$CDM cosmology from
\cite{PlanckCosmo} as our fiducial background cosmology.
The parameter set we use is $h=0.6774$, $\Omega_b h^2 = 0.0223$,
$\Omega_c h^2 = 0.1188$, $n_s=0.9667$, and $\sigma_8 = 0.8159$.
We use this fiducial cosmology to convert observed quasar coordinates
(right ascension, declination, and redshift) to Cartesian coordinates
during the estimation of the power spectrum of the sample
(see \S\ref{sec:power-estimation}). The fiducial
cosmology also determines the shape of the real-space matter power spectrum,
which is used in our theoretical modeling (see \S\ref{sec:model}).

\subsection{Power spectrum estimation}\label{sec:power-estimation}

We begin by defining the weighted quasar density fields \cite{Feldman:1994}
\begin{equation}\label{eq:Fr}
  \tilde{F}(\vr) =\tilde{w}_{\rm{tot}} \left[n'_\qso(\vr) - \synalpha n_\syn(\vr)\right] \quad,\quad F_\ell(\vr) = w_{\rm{tot},\ell} \left[n'_\qso(\vr) - \synalpha n_\syn(\vr) \right],
\end{equation}
where $n'_\qso$ and $n_\syn$ are the number densities of the quasar sample and
a synthetic catalog of random objects, respectively. The total weights are the product of FKP and the optimal redshift weights
\begin{align}
\tilde{w}_{\rm{tot}}(z) = w_{\rm FKP}(z) \tilde{w}(z)\quad,\quad w_{\rm{tot},\ell}(z) = w_{\rm FKP}(z) w_\ell(z) \;,
\end{align}
and they are applied to both the quasar and synthetic samples.
The synthetic catalog contains unclustered objects -- it is
used to define the expected mean density of the survey, accounting for the
radial and angular selection functions. The factor $\synalpha$ gives the
ratio of quasars to synthetic objects and serves to properly normalize the
number density of the synthetic catalog.
In our notation, quantities marked with a prime ($\prime$)
include the completeness weights $\wc$ specified in Section~\ref{sec:comp-weights}.
The synthetic catalog defines our expected number density, and as such, does
not require completeness weights. The synthetic sample has a number density
$1/\synalpha$ times more dense than the true sample. We assume that, on average,
the relation $\la  n'_\qso(\vr) \ra = \synalpha \la n_s(\vr) \ra$ holds true.
We define $\synalpha$ as
$\synalpha= N'_\qso / N_\syn$, where $N'_\qso = \sum_{\qso} w_c$
and $N_\syn$ is the total number of objects in the synthetic catalog.

Now, the multipoles of the cross-correlation between
the weighted density fields can be estimated
following \cite{Yamamoto:2006},

\begin{equation}\label{eq:Pell-yama}
\hat{P}_\ell = \frac{2\ell+1}{A_\ell} \int \frac{d\Omega_k}{4\pi}
									\left [
									\int d\vrone \ \tilde{F}(\vrone) e^{i \vk \cdot \vrone}
                                    \int d\vrtwo \ F_\ell(\vrtwo)
                                    e^{-i \vk \cdot \vrtwo}
                                    \Leg_\ell(\vkhat \cdot \vrhat_2)
                                    \right ]  - S_\ell,
\end{equation}
where we have introduced the shot noise contribution $S_\ell$, defined as

\begin{equation}\label{eq:S-integral}
S_\ell = A_\ell^{-1} \int d\vr \ n'_\qso(\vr) (\wc(\vr) + \synalpha) \tilde{w}_{\rm{tot}}w_{\rm{tot},\ell}(\vr) \Leg_\ell (\vkhat \cdot \vrhat),
\end{equation}
which is different than zero only for the monopole $\ell=0$.
The normalization is  defined as
\begin{equation}\label{eq:A-integral}
 A_\ell = \int d\vr\; w_{\rm{tot},\ell}(\vr)\tilde{w}_{\rm{tot}}(\vr) [n'_\qso(\vr)]^2\;.
\end{equation}
We compute the shot noise (equation~\ref{eq:S-integral}) and the normalization (equation~\ref{eq:A-integral}) as discrete sums over the
quasar and synthetic catalogs. To do so, we make use of the following relation:

\begin{equation}
  \label{eq:int-to-sum}
  \int d\vr \; n'_\qso(\vr) \ldots  \longrightarrow \sum_i^{N_\qso} \wc(\vri) \ldots
    \longrightarrow \synalpha \sum_i^{N_\syn} \ldots ,
\end{equation}
where the integral can be expressed equivalently as a sum over the quasar
or synthetic catalogs. Thus, the normalization $A_\ell$ can be computed as

\begin{align}
  \label{eq:norm.data}
  A_\ell & = \sum_i^{N_\qso} n'_\qso (\vri) w_c(\vri) w_{\rm{tot},\ell}(\vri)\tilde{w}_{\rm{tot}}(\vri) \\
  \label{eq:norm.randoms}
    &  = \synalpha \sum_i^{N_\syn} n_s (\vri) w_{\rm{tot},\ell}(\vri)\tilde{w}_{\rm{tot}}(\vri)\;.
\end{align}
Note that while equations~\ref{eq:norm.data} and \ref{eq:norm.randoms}
are equivalent on average, in practice, we use the latter equation to estimate the normalization
due to the increased number density of the synthetic catalog.
Similarly, we can express the shot noise contribution to the monopole
(equation~\ref{eq:S-integral}) as

\begin{equation}
  \label{eq:shotnoise}
  S_0 = A_0^{-1} \left[ \sum_i^{N_\qso} \wc^2(\vri)  \tilde{w}_{\rm{tot}}(\vri)w_{\rm{tot},0}(\vri) +
        \alpha^{\prime 2}_\syn \sum_i^{N_\syn}  \tilde{w}_{\rm{tot}}w_{\rm{tot},0}(\vri) \right],
\end{equation}
where the two terms compute the contributions to the shot noise from the
quasar and synthetic catalogs, respectively. There is some uncertainty
surrounding the impact of fiber collisions and completeness weights on the
Poisson shot noise calculation of equation~\ref{eq:shotnoise}
\cite{Beutler:2014, Beutler:2017, Grieb:2017}. We choose to use the
standard Poisson expression and vary a shot noise parameter while
performing parameter estimation to account for any discrepancies
(see section~\ref{sec:model}).

Our implementation of equation~\ref{eq:Pell-yama} uses the FFT-based
estimator of \cite{Hand:2017}. This estimator builds upon similar estimators
presented in \cite{Bianchi:2015,Scoccimarro:2015}, but reduces the number
of FFTs required per multipole using a spherical harmonic decomposition.
We calculate the power spectrum multipoles as 

\begin{equation}\label{eq:Pell-FFT-estimator} 
\hat{P}_\ell(k) = \frac{2\ell+1}{A_\ell} \int \frac{d\Omega_k}{4\pi} \tilde{F}(\vk) F_\ell(-\vk),
\end{equation}
with

\begin{align}\label{eq:Fell}
F_\ell(\vk) & \equiv \int d\vr \ F_\ell(\vr) e^{i \vk \cdot \vr}
								\Leg_\ell(\vkhat \cdot \vrhat), \nn \\
			& = \frac{4\pi}{2\ell+1} \sum_{m=-\ell}^{\ell}
            				\Ylm(\vkhat) \int d\vr \ F_\ell(\vr) \Ylm^*(\vrhat) e^{i \vk \cdot \vr},
\end{align}
where $\Ylm$ are spherical harmonics. Note that equation~\ref{eq:Fell}
requires the calculation of $2\ell+1$ FFTs for a multipole of order $\ell$.

To compute the FFTs required by our estimator, we
estimate the overdensity field on a mesh of $1024^3$ cells
for the quasar and synthetic catalogs using a Triangular Shaped Cloud (TSC)
interpolation scheme (see e.g., \cite{Hockney:1981}).
When interpolating to the mesh, each quasar contributes a weight of
$\wc w_{\rm{tot}}$ and each synthetic object a weight of $w_{\rm{tot}}$.
When computing FKP weights, we use a fiducial power
spectrum value of $P_0 = 3 \times 10^{4} \hMpcc$, roughly equal to the
expected power on the scales where PNG is prominent in our sample, $k \simeq 0.03 \ihMpc$.
We use the interlaced grid technique of \cite{Sefusatti:2016,Hockney:1981} to
limit the effects of aliasing, and we correct for any artifacts of the
TSC scheme using the correction factor of \cite{Jing:2005}.
With the combination of TSC interpolation and interlacing, we are able
to measure the power spectrum multipoles up to $k = 0.4 \ihMpc$ with
fractional errors at the level of $10^{-3}$ \cite{Sefusatti:2016}.
To perform these operations, as well as estimate the power spectrum
multipoles via equation~\ref{eq:Pell-FFT-estimator}, we utilize
the massively parallel implementations available as part of the open-source
Python toolkit \texttt{nbodykit} \cite{Hand:2017b}.

\subsection{Modeling}\label{sec:model}

\subsubsection{The power spectrum model}

We use linear theory to predict the quasar power spectrum in redshift-space
\cite{Kaiser:1987}

\begin{equation}\label{eq:Pqso-model}
    P_\qso(k, \mu) = G(k,\mu; \sigma_P)^2 \; \left[ \btot(k) + f \mu^2 \right]^2 P_m(k) + N,
\end{equation}
where $P_m$ is the real-space matter power spectrum, $N$ is a free parameter
accounting for residual shot noise, and $\btot$ is the total
quasar bias, including PNG, given by

\begin{equation}
\label{eq:btot}
  \btot(k) = b_\qso + \Delta b = b_\qso + \fnl (b_\qso - p)\atilde(k),
\end{equation}
where $b_\qso$ is the linear bias of the quasar sample, and
$\atilde$ is defined in \eq{eq:atilde}. To account for
redshift-space related damping of the power spectrum,
we include a Lorentzian damping function,

\begin{equation}
  G(k,\mu; \sigma_P) = \left[1 + (k\mu\sigma_P)^2/2 \right]^{-1},
\end{equation}
with a single free parameter $\sigma_P$, which represents the typical
damping velocity dispersion. The physical motivation for the inclusion of
$G(k,\mu)$ is the Finger-of-God (FOG) effect in redshift space due to the
virial motions of the quasar within its host dark matter halo \cite{Jackson:1972}.
However, the damping term also accounts for errors in the spectroscopic
redshift determination of the quasars \cite{Dawson:2016}. The effect can
be estimated for the DR14Q sample as $\sigma_z = 300 \; \mathrm{km \; s^{-1}}$
for $z < 1.5$ and $\sigma_z = [400 \cdot (z - 1.5) + 300] \; \mathrm{km \; s^{-1}}$
for $z > 1.5$ \citep{Dawson:2016}.

The multipoles of the power spectrum are then computed as

\begin{equation}\label{eq:pole-model}
P_{\ell,\qso}(k) = \frac{2\ell + 1}{2} \int_{-1}^{1} d\mu P_\qso(k,\mu) \mathcal{L}_\ell(\mu),
\end{equation}
We evaluate the linear, real-space matter power spectrum $P_m(k)$ and the
transfer function in $\Delta b$ using
the \texttt{classylss} software \cite{classylss}, which provides Python
bindings of the \texttt{CLASS} CMB Boltzmann solver \cite{Blas:2011}.
We evaluate the linear power spectrum
using the fiducial cosmology and keep the shape of the power spectrum fixed during parameter estimation. This choice
assumes that the uncertainty as determined by \cite{Planck:2016}
for most of the parameters which define the shape of the power spectrum
is much smaller than the uncertainty of our measurement and
can be neglected. This has been shown to be a reasonable assumption for
current data sets, e.g., \cite{Beutler:2014,Gil-Marin:2016}.

\subsubsection{Window function and effective redshift}
\label{sec:zeff}
The measured power spectrum multipoles are the result of the convolution of the true underlying power spectrum with the Fourier transform a window function $W(\mathbf{s})$, defined by the footprint on the sky and the redshift selection function. It is easy to see that the ensemble average of the estimator in \eq{eq:Pell-yama} measures the following multipoles of the power spectrum $P_{A,eff}(k)$ \cite{Wilson:2017,CastorinaWhite2018a,CastorinaWhite2018b,BeutlerCastorina}
\begin{align}
\label{eq:PAeff}
P_{A,eff}(k) &= (2A+1)\int \frac{d \Omega_k}{4\pi} \,\mathrm{d}^3 s_1 \,\mathrm{d}^3 s_2 \,e^{i \vec{k} (\vec{s}_2-\vec{s}_1)} \delta(\vec{s}_1) \delta(\vec{s}_2)  W(\vec{s}_1)W(\vec{s}_2)\mathcal{L}_A(\hat{k}\cdot \hat{s}_1) \\
& =(-i)^A (2A+1) \sum_{\ell,\, L}\tj{\ell}{L}{A}{0}{0}{0}^2\int \mathrm{d}s\,s^2 j_A (ks)\,\int \mathrm{d}s_1\,s_1^2\,  \xi_\ell(s;s_1(z)) Q_L(s;s_1(z))
\end{align}
where we have defined the multipoles of the window function
\begin{align}
Q_L(s) &\equiv (2L+1)\int \mathrm{d} \Omega_s \int \mathrm{d}^3 s_1  W(\vec{s}_1)W(\vec{s}+\vec{s}_1)\mathcal{L}_L(\hat{s}\cdot \hat{s}_1) \\
&\equiv\int \mathrm{d} s_1 \ s_1^2 \,Q_L(s;s_1)\,.
\end{align}
In the above equations we have chosen the direction to one of the two QSOs to be the line of sight for the multipole decomposition, $s_1(z)$, explicitly accounting for the redshift evolution of the signal.
In principle, one should perform the redshift integral $\int \mathrm{d} \,s_1(z)$ in \eq{eq:PAeff} for each evaluation of the model parameters in the likelihood, which makes the data analysis numerically quite challenging.
However, as we will now show, an effective redshift approximation is often very close to the full answer in \eq{eq:PAeff}.

\begin{figure}
\centering
\includegraphics[width=0.475\textwidth]{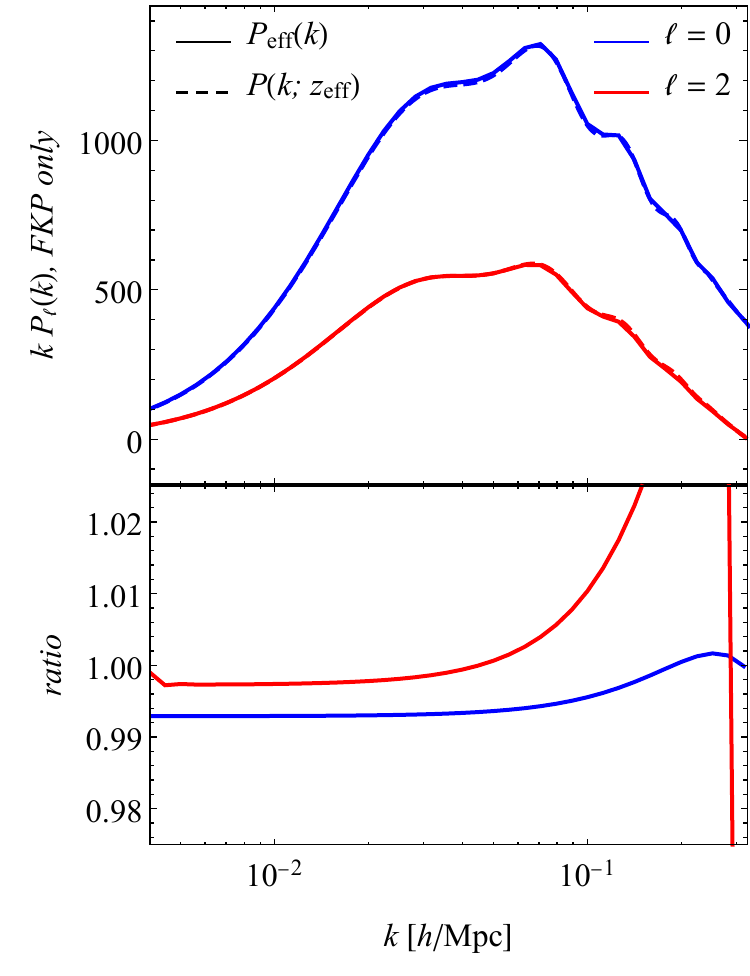}
\includegraphics[width=0.475\textwidth]{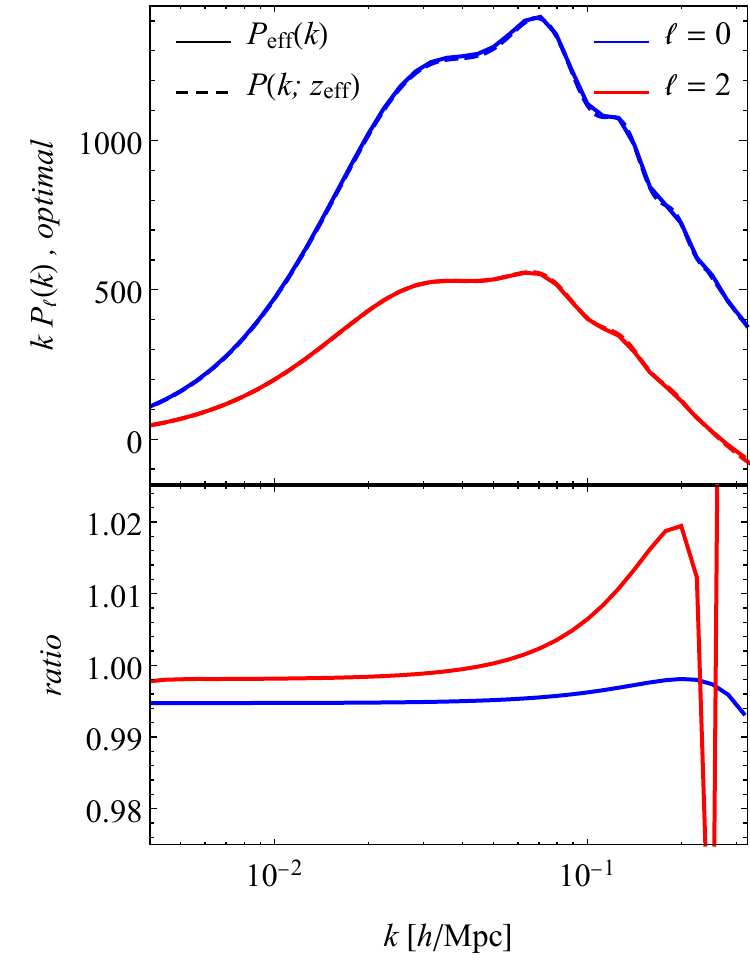}
\caption{Accuracy of the effective redshift approximation for the eBOSS DR14 QSOs sample (NGC). \emph{Left panel}: The monopole, blue lines, and quadropole, red lines, evaluated at the effective redshift defined by FKP weights, dashed lines, compared to the full integral of the signal power spectrum over the QSOs selection function, continuous lines. The accuracy for both the monopole and quadrupole is at a percent level, well withing the error in the measurements.
\emph{Right panel}: Same as the left panel but including $\fnl$ optimal weights. The amplitude and shape of the multipoles of the power spectrum have changed compared to the plot on the left, but new effective redshift can be defined to provide an excellent description of the full redshift integral.
}
\label{fig:zeff}
\end{figure}

Under a generic set of weights $w(z)$ of the density field $\delta$, the effective redshift is defined as 
\begin{align}
z_{\rm eff} = \frac{\int  dz\ n(z)^2[\chi^2/H(z)]\ w(z)^2 z}
       {\int dz\ n(z)^2[\chi^2/H(z)]w(z)^2}
 \end{align}
and the power spectrum evaluated at the effective redshift reads
\begin{align}
\label{eq:Pkzeff}
P_A(k;z_{\text{\text{eff}}}) =(-i)^A (2A+1) \sum_{\ell,\, L}\tj{\ell}{L}{A}{0}{0}{0}^2\int \mathrm{d}s\,s^2 j_A (ks) \xi_\ell(s;z_{\text{eff}}) Q_L(s)
\end{align}
For FKP weights and smoothly varying selection function the above expression is sub-percent accurate, even in large redshift bins \citep{Vlah2016}. 
In particular, on scales where linear theory is a good approximation, it is always possible to define an effective redshift because in linear theory redshift evolution preserves the shape of the power spectrum.
For the DR14 QSOs in NGC, neglecting for the moment the angular mask, assuming the fiducial cosmological parameters and the model in \eq{eq:Pqso-model}, the accuracy of the effective redshift approximation is shown in the left hand panel of \fig{fig:zeff} for FKP weights. For NGC, between $0.8\le z \le2.2$ the effective redshift is $z_{\rm eff} = 1.52$.
The measured monopole power spectrum, in blue, can be accurately described by a model evaluated at the effective redshift of the survey, with deviations smaller than 1\%. Moreover the difference between \eq{eq:Pkzeff} and \eq{eq:PAeff} is well captured by a constant, which can therefore absorbed when marginalizing over galaxy bias.
Similar conclusions hold for the quadrupole. The large difference at high $k$ is a mere consequence of the fact that the quadropole crosses zero at $k\simeq 0.25\,\kMpc$, the relative deviation is still quite small. The bottom line is that, within FKP weighting, the error introduced by evaluating the model at an effective redshift is always well below the measurement uncertainties.

Our optimal weights are smooth in redshift, thus we expect to be able to evaluate the theoretical model at new effective redshifts, one for the monopole
\begin{align}
z_{0,\rm eff} = \frac{\int dz\ n(z)^2[\chi^2/H(z)]\ w_{FKP}^2 \tilde{w}(z) w_0(z)z}
       {\int dz\ n(z)^2[\chi^2/H(z)]w_{FKP}^2\tilde{w}(z) w_0(z)}\;,
\end{align}
and one for the quadrupole
\begin{align}
z_{2,\rm eff} = \frac{\int dz\ n(z)^2[\chi^2/H(z)]\ w_{FKP}^2\tilde{w}(z) w_2(z)z}
       {\int dz\ n(z)^2[\chi^2/H(z)] w_{FKP}^2\tilde{w}(z) w_2(z)}\;.
\end{align}
Since the optimal analysis up-weights high-redshift galaxies where the PNG signal is larger, the effective redshift goes up compared to the FKP-only case, $z_{\rm eff} = 1.52$, to
\begin{align}
z_{0,\rm eff}=1.64\;,\;z_{2,\rm eff}=1.58\;,
\end{align}
for  $p=1.0$, and to 
\begin{align}
\label{eq:zell_eff}
z_{0,\rm eff}=1.74\;,\;z_{2,\rm eff}=1.70\;,
\end{align}
for $p=1.6$.
For an optimally weighted case the comparison between the effective redshift approximation, Eq.~\ref{eq:Pkzeff}, and the full integral over redshift, Eq.~\ref{eq:PAeff}, is shown in the right hand panel of \fig{fig:zeff}. We find that despite the large redshift range covered by eBOSS QSOs, evaluating the power spectrum at the effective redshift is accurate at the percent level for both the monopole and the quadropole.
We have further tested this assumptions in mock catalogs, finding that we can reproduce the weighted clustering using the effective redshifts defined above.

We finally stress that effective redshifts can be defined using optimal weights with respect to any other cosmological parameter, and thus, upon checking they provide an accurate description of the data, vastly simplify the cosmological analysis.
For instance, see \citep{DR14Zhao18BAO} for an implementation of the effective redshift approximation for BAO and RSD in eBOSS data.

\begin{figure}[tb]
\centering
\includegraphics[width=0.475\textwidth]{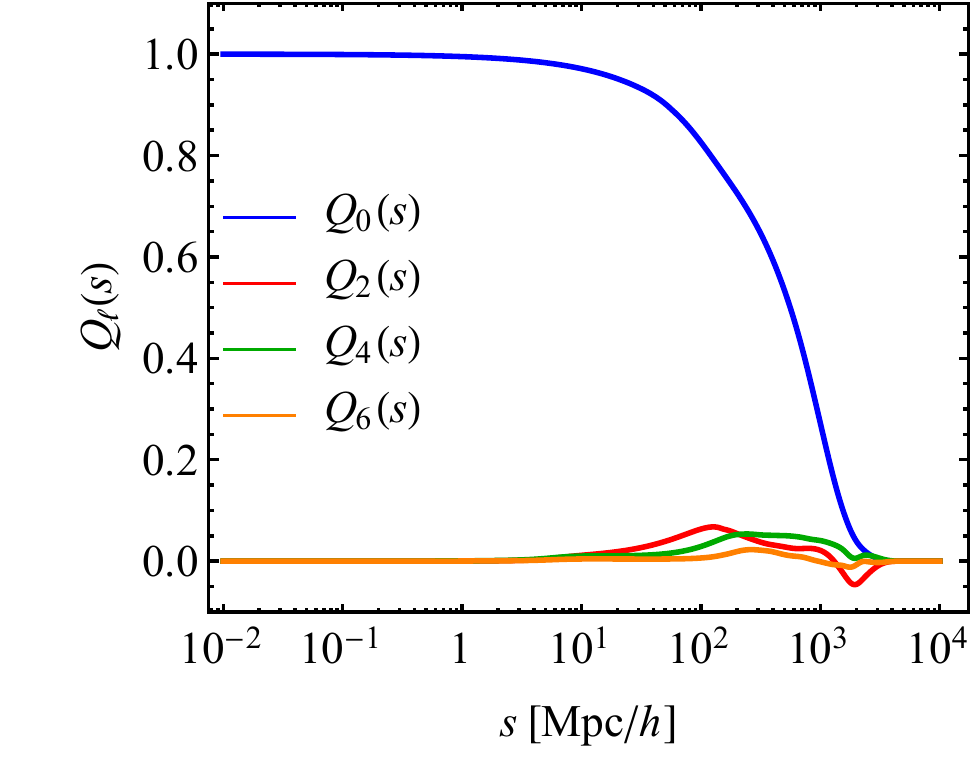}
\includegraphics[width=0.475\textwidth]{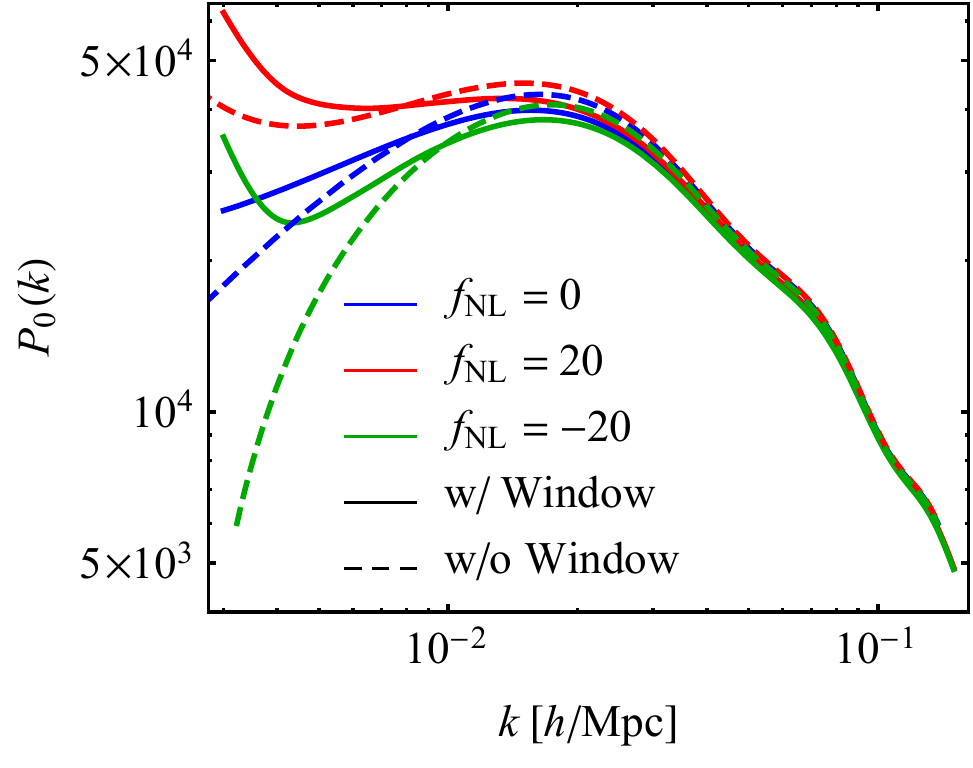}
\caption{The window function multipoles in configuration space (left)
and the effects of the window function on linear Kaiser power
spectrum multipoles (right) for the eBOSS DR14Q NGC survey geometry.
In the right panel, the solid lines show the original multipoles, while
the dashed lines correspond to the model after convolution
with the window function, $\hat{P}_\ell(k)$. The main consequence of the presence 
of the survey geometry is a change in power on large scales.}
\label{fig:window-effects}
\end{figure}

The shape of the window function multipoles $Q_L(s)$ and their effect on the monopole of the eBOSS QSOs power spectrum at $z_{\rm eff}$ is shown in Figure~\ref{fig:window-effects}. The main effect of the survey mask is to reduce the amplitude of clustering at large scales compared to the true underlying power spectrum, and therefore needs to be properly taken into account for unbiased estimates of PNG.
Given the high effective redshift and the small fraction of the sky covered by DR14Q, wide angle effects in redshift space distortions and their possible coupling to the survey mask can be safely neglected \cite{CastorinaWhite2018a,BeutlerCastorina}.

\subsection{Parameter estimation}\label{sec:param-estimation}

In summary, our base model has three free parameters in each patch of the sky: the linear bias
$b_1$, the damping velocity dispersion $\sigma_P$, and the
shot noise parameter $N$. When fitting for PNG, we introduce $\fnl$ as
an additional free parameter. 
Since we are not interested in BAO information or RSD we keep cosmological parameters fixed to Planck best fit values and do not include the Alcock-Paczynski (AP) effect
\cite{Alcock:1979}. Actually, the onset of the PNG signal is fixed by the position of the wavenumber associated with the size of the horizon at matter-radiation equality, which is one of the most precise derived parameters measured by Planck \cite{PlanckLegacy,PlanckCosmo}.
Since we have shown that the optimal quadratic estimator, in the limit of diagonal pixel covariance, boils down to a standard measurement of the power spectrum, we don't need to remove the any estimator bias and just model the expected signal. 
We estimate the best-fit parameters of the model using a likelihood analysis.
We assume that the probability that our data vector $\vD$ corresponds
to a realization of our model $\vT(\vtheta)$ is given by a
multi-variate Gaussian of the form,

\begin{equation}\label{eq:likelihood}
  \mathcal{L}(\vD|\vtheta, \vPhi) \propto \exp \left [-\frac{1}{2}\chi^2(\vD,\vtheta,\vPhi)\right],
\end{equation}
where $\vtheta$ is our vector of model parameters, and
$\chi^2$ takes the quadratic form,

\begin{equation}
  \chi^2(\theta) = \sum_{ij} (D_i - T_i(\vtheta)) \Phi_{ij} (D_j - T_j(\vtheta)),
\end{equation}
and $\vPhi$ is the inverse of the covariance matrix $\vC$, often referred as
the precision matrix.

When performing our likelihood analysis, our data vector $\vD$ consists
of the monopole and quadrupole, measured using the procedure outlined in
section~\ref{sec:power-estimation}. We use linearly spaced bins of
width $\Delta k = 0.001 \ihMpc$. With the first bin separation at
$k \sim 0.005 \ihMpc$ and extending to $\kmax = 0.3 \ihMpc$, we have a total
of 120 data points in $\vD$ (60 bins per multipole).

We estimate the covariance matrix of our data measurement using the 1000 EZ mock
realizations, described previously in section~\ref{sec:ezmocks}.
As the covariance is computed from a finite number of mock realizations, its
inverse $\vPhi$ provides a biased estimate of the true precision matrix
due to the skewed nature of the inverse Wishart distribution \cite{Hartlap:2007}.
To correct for this bias, we re-scale the precision matrix as

\begin{equation}
  \vPhi' = \frac{N_\mathrm{mock} - n_b - 2}{N_\mathrm{mock}-1} \vPhi.
\end{equation}
We performing our likelihood analysis following equation~\ref{eq:likelihood},
we use the rescaled precision matrix $\vPhi'$. In our analysis,
we use $N_\mathrm{mocks} = 1000$ and $n_b = 120$, yielding a Hartlap factor
of $\sim$0.88. Following \citep{DR14Gil-Marin18RSD} we do not include the extra correction of \cite{Percival:2014} since it has a minor impact on the errors.

We find the best-fitting model parameters using the LBFGS nonlinear
minimization algorithm \cite{Byrd:1995}. We verify that the
minimization procedure converges by starting the algorithm from a number
of different initialization states. We compute the full posterior
distribution of the parameters of interest using the \texttt{emcee}
software \cite{Foreman-Mackey:2013} to perform Markov chain Monte Carlo
(MCMC) sampling. We assume broad, uniform priors on all parameters of
interest such that the priors serve only to bound the parameter values to
the largest possible physically meaningful parameter space; they do not have
an impact on our derived posterior distributions.

\section{Fisher Information}
\label{sec:Fisher}
It is useful to look at what a Fisher analysis \citep{Tegmark:1998} based on eBOSS number densities and sky area returns for the error on $\fnl$ using measurements of scale-dependent bias.
This will tell us the best possible constraints, and on how much improvement we can expect from an optimal analysis. 

For simplicity we will assume shot-noise and the redshift error $\sigma_v$ are perfectly known, as they both have a small impact on the final $\fnl$ bounds. Since we are interested in quantifying the maximum information of the survey we assume QSOs in NGC and SGC have the same value of linear bias. 
We consider measurements of the monopole and quadropole only, for which the Fisher matrix reads
\begin{align}
\label{eq:Fij}
%F^{\rm{eff}}_{ij} = \frac{1}{2} V \int_{k_{\rm min}}^{k_{\rm max}} \frac{\mathrm{d}^3 k}{(2\pi)^3} n_\qso^2(z_{\rm eff})\frac{\partial P_\qso(k,\mu;z_{\rm eff}) / \partial \theta_i}{n_\qso(z_{\rm eff})P_\qso(k,\mu;z_{\rm eff})+1} \frac{\partial P_\qso(k,\mu;z_{\rm eff}) / \partial \theta_j}{n_\qso(z_{\rm eff})P_\qso(k,\mu;z_{\rm eff})+1} 
F_{ij} =  V \int_{k_{\rm min}}^{k_{\rm max}}\frac{\mathrm{d} k\,k^2}{2\pi^2}\frac{\partial\vb{P}(k)}{\partial\theta_i}^{\rm T} \mathcal{\mathbf{C}}(k)^{-1} \frac{\partial\vb{P}(k)}{\partial\theta_j}\,
%\frac{1}{2} f_{\rm sky}  &\int_{z_{\rm min}}^{z_{\rm max}} \, \mathrm{d} z\, 4 \pi \frac{\chi(z)^2}{H(z)} n^2_\qso(z)   \notag \\ \,\times
%& \int_{k_{\rm min}}^{k_{\rm max}}\frac{\mathrm{d}^3 k}{(2\pi)^3} \frac{\partial P_\qso(k,\mu;z_{\rm eff}) / \partial \theta_i}{n_\qso(z)P_\qso(k,\mu;z)+1} \frac{\partial P_\qso(k,\mu;z) / \partial \theta_j}{n_\qso(z)P_\qso(k,\mu;z)+1} \;,
\end{align}
where $\theta = \{b_\qso,\fnl\}$, $\vb{P}(k)=\{P_0(k),P_2(k)\}$ is a vector formed by the monopole and quadrupole of the power spectrum, and $\mathcal{\mathbf{C}}(k)^{-1}$ is the inverse of the covariance matrix of the measurements. The fiducial value for the QSO bias is taken from the fitting function in \citep{Laurent:2017}, while the fiducial redshift error from \cite{Dawson:2016}.

The Fisher matrix can be evaluated at the effective redshift defined by FKP weights or by the optimal ones.
%\begin{align}
%\label{eq:Fij}
%F_{ij} =  \frac{1}{2} f_{\rm sky}  &\int_{z_{\rm min}}^{z_{\rm max}} \, \mathrm{d} z\, 4 \pi \frac{\chi(z)^2}{H(z)} n^2_\qso(z)   \notag \\ \,\times
%& \int_{k_{\rm min}}^{k_{\rm max}}\frac{\mathrm{d}^3 k}{(2\pi)^3} \frac{\partial P_\qso(k,\mu;z) / \partial \theta_i}{n_\qso(z)P_\qso(k,\mu;z)+1} \frac{\partial P_\qso(k,\mu;z) / \partial \theta_j}{n_\qso(z)P_\qso(k,\mu;z)+1} \;.
%\end{align}
In both cases $\sigma_{\fnl}$ is defined via $(\sqrt{F^{-1}})_{22}$.
The DR14 catalog contains QSOs from redshift $0.5<z<3.5$, of which the redshift range $0.8<z<2.2$ corresponds to the fiducial survey. We will repeat the calculation in both redshift ranges, to assess the gains of an extended, in redshift, analysis.
We furthermore distinguish between NGC and SGC, and thus perform two separate Fisher calculations which are then added together.
NGC covers a larger area of the sky than SGC, by $\simeq30$\%, and will therefore be more constraining. 
As discussed in Section \ref{sec:theory}, Eqs. \ref{eq:def_bias} and \ref{eq:btot}, the response of a generic tracer to the presence of PNG depends on one parameter $p$, that for QSOs takes a value between 1 and 1.6. We therefore repeat the Fisher matrix calculation, and in Section \ref{sec:constraints} the fit to the data, for both values of $p=1,1.6$. 

\subsection{Fiducial Survey: $0.8<z<2.2$}
\begin{figure}[tb]
\centering
\includegraphics[width=0.475\textwidth]{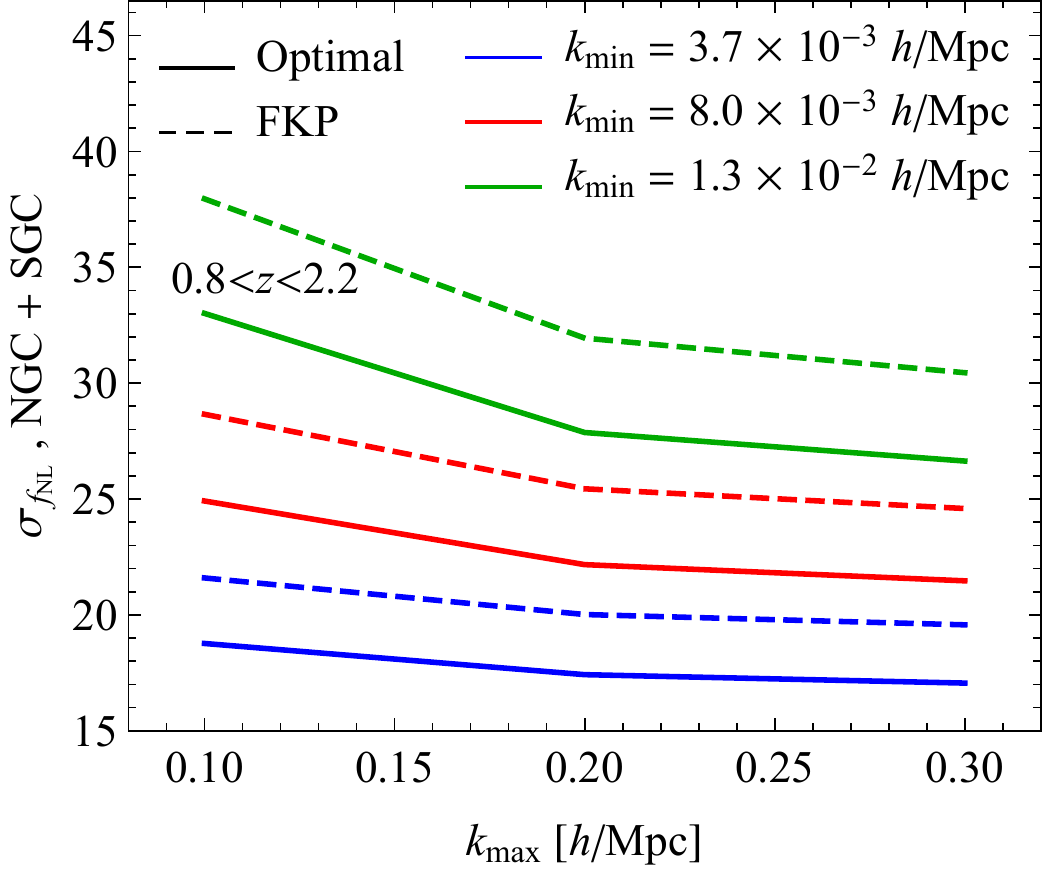}
\hfill
\includegraphics[width=0.475\textwidth]{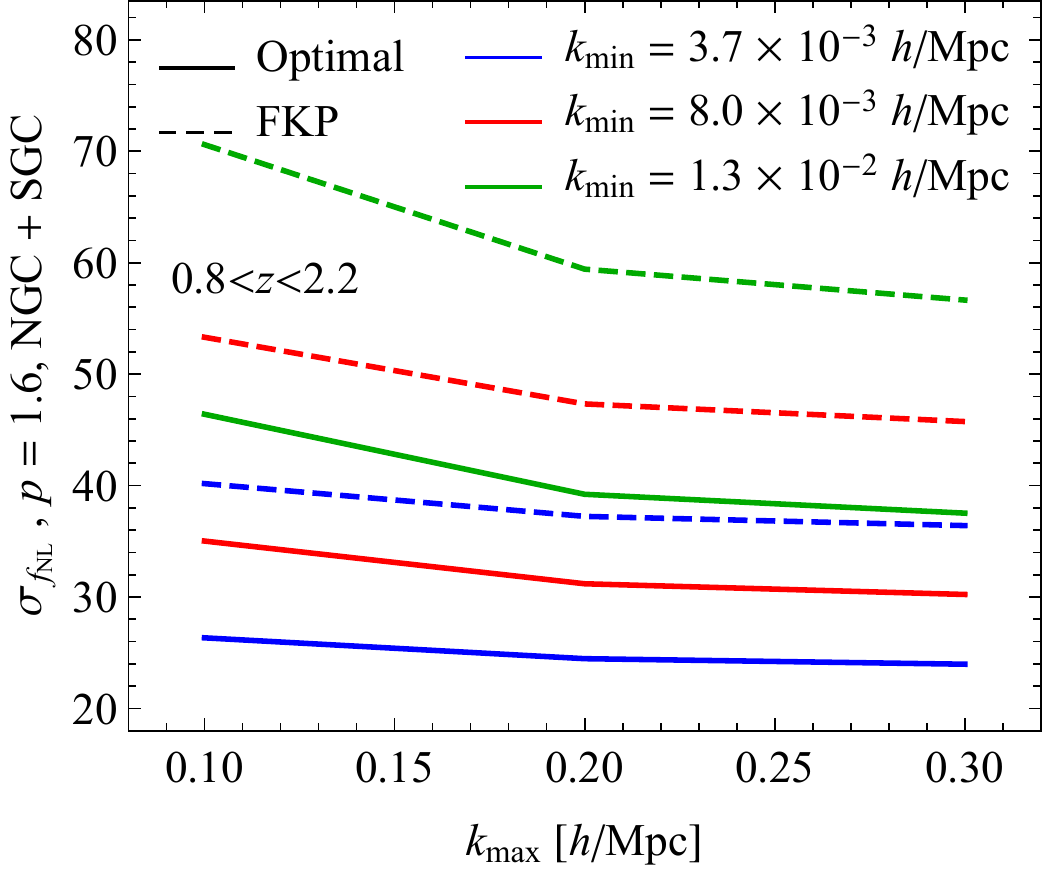}
\caption{Constraints on PNG from a combined analysis of NGC and SGC in the redshift range $0.8<z<2.2$. Different colors corresponds to different value of $k_{\rm min}$. Dashed lines show the FKP only weights, whereas continuous lines refer to the optimal analysis. Left and right panels correspond to $p=1.0$ and $p=1.6$ respectively.}
\label{fig:Baseline}
\end{figure}

The left panel of Figure~\ref{fig:Baseline} shows the constraints on $\fnl$, for $p=1$, in NGC plus SGC, as a function of $k_{\rm max}$ and for different values of $k_{\rm min}$. The values of $k_{\rm min}$ correspond to the first three $k$-bins of the measured power spectra.
The dashed lines correspond to the FKP weighting, while the continuous one to the optimal analysis.
The error $\sigma_{\fnl}$ strongly depends on the largest scales included in the analysis, since the signal peaks at low $k$, but it very weakly changes with $k_{\rm max}$.
Comparing the standard analysis to the optimal one, we find that the optimal method provides roughly 20\% better error bars than the FKP one, with larger improvement for higher values of $k_{\rm min}$. 
The right panel in Figure~\ref{fig:Baseline} displays the results of the Fisher analysis for $p=1.6$. The optimal analysis for $p=1.6$ moves the effective redshift further up, in order to compensate for the reduced response to $\fnl$. We thus expect, compared to $p=1$, larger difference with respect to the FKP-only weighting. We indeed find 40-60\% benefit of the optimal analysis comparing it to a standard one. 
A comparison of the two panels in Figure~\ref{fig:Baseline} also shows that $\sigma_{\fnl}$ degrades by almost a factor of 2 going from $p=1$ to $p=1.6$ in the standard FKP analysis, whereas with our method we lose only 50\% of the constraining power. 
% \begin{figure}[tb]
% \centering
% \includegraphics[width=0.475\textwidth]{fig_zeff.pdf}
% \hfill
% \includegraphics[width=0.475\textwidth]{fig_optimal.pdf}
% \caption{Fisher forecast on $\sigma_{\fnl}$, assuming $p=1.0$ and the fiducial redshift range $0.8<z<2.2$, as a function of the maximum scale included in the analysis, $k_{\rm max}$. Different colors corresponds to different value of $k_{\rm min}$, while different dashing to the North, continuous, and South, dashed, Galactic cups.}
% \label{fig:Baseline_p1p0}
% \end{figure}

% \begin{figure}[tb]
% \centering
% \includegraphics[width=0.475\textwidth]{fig_zeff_p1p6.pdf}
% \hfill
% \includegraphics[width=0.475\textwidth]{fig_optimal_p1p6.pdf}
% \caption{Same as \fig{fig:Baseline_p1p0} for $p=1.6$. }
% \label{fig:Baseline_p1p6}
% \end{figure}

It is worth emphasizing again that we do not know the exact value of $p$ for the eBOSS QSOs sample. This implies that an analysis of the eBOSS data assuming $p=1$ or $p=1.6$ will not necessarily return the error bar of the Fisher calculation described above.
It is nonetheless reasonable to expect that, in absence of systematic effects at low-$k$, $\sigma_{\fnl}\lesssim20$ and $\sigma_{\fnl}\lesssim30$, for $p=1$ and $p=1.6$ respectively, are in the reach of eBOSS DR14Q data.
% Figure~\ref{fig:Baseline} presents the total constraining power of the eBOSS sample on $\fnl$, summing up NGC and SGC.  This represents the best possible constraints on PNG achievable using scale dependent bias in a measurement of the power spectrum in the redshift range $0.8<z<2.2$.
% We find that optimal analysis yields 25-30\% improvement over standard methods for $p=1$, and about 40-60\% for $p=1.6$. In absence of systematic effects at low-$k$ we conclude that $\sigma_{\fnl}\lesssim20$ and $\sigma_{\fnl}\lesssim30$, for $p=1$ and $p=1.6$ respectively, are in the reach of eBOSS data.

\subsection{Including QSOs at $z>2.2$}

The analysis of the previous section suggests that extending the $\fnl$ analysis to $z>2.2$ could significantly increase the sensitivity to PNG. The benefit of a larger redshift coverage is two-fold. First, at fixed $k_{\rm min}$, sample variance is reduced in a larger volume simply because more modes are available. The error $\sigma_{\fnl}$ indeed roughly scales with $V^{1/2}$ at fixed $k_{\rm min}$, see \eq{eq:Fij}. Second, since the signal peaks at the largest scales, including lower $k$ modes into the analysis shrinks the error bars by another factor of $V^{1/6}$. The latter improvement would however require a careful study of the systematic effects at large scale, as described in Section \ref{sec:spec-weights}. 
In this section we calculate the Fisher information of the full redshift range covered by the eBOSS survey, providing a motivation to further investigate and reduce the systematics at low $k$.
The redshift distribution of eBOSS QSOs is such that at $z>2.5$ the number of objects drops very quickly. In terms of the effective redshift defined in Section~\ref{sec:zeff} we find
\begin{align}
0.8<z<2.5\;:\quad z_{\rm eff} = 1.54\;,\,
\begin{cases}
z_{\rm{eff},0} = 1.72\;,\;z_{\rm{eff},2} = 1.66\,, & \text{for } p=1 \\
z_{\rm{eff},0} = 1.83\;,\;z_{\rm{eff},2} = 1.78\,, & \text{for } p=1.6
\end{cases}
\end{align}
and
\begin{align}
0.8<z<3.5\;:\quad z_{\rm eff} = 1.56\;,\,
\begin{cases}
z_{\rm{eff},0} = 1.78\;,\;z_{\rm{eff},2} = 1.69\,, & \text{for } p=1 \\
z_{\rm{eff},0} = 1.89\;,\;z_{\rm{eff},2} = 1.82\,, & \text{for } p=1.6\;.
\end{cases}
\end{align}
The inclusion of QSOs in the range $2.2<z<2.5$ basically does not change the FKP effective redshift, but it moves up the optimal effective redshift by a substantial amount compared to \eq{eq:zell_eff}. This indicates that a large amount of signal could become accessible by including $z<2.5$ QSOs in the analysis. Minor shifts in the effective redshift are produced by adding QSO all the way to $z=3.5$, suggesting that this redshift range will likely only help to reduce sample variance.

\begin{figure}[tb]
\centering
\includegraphics[width=0.49\textwidth]{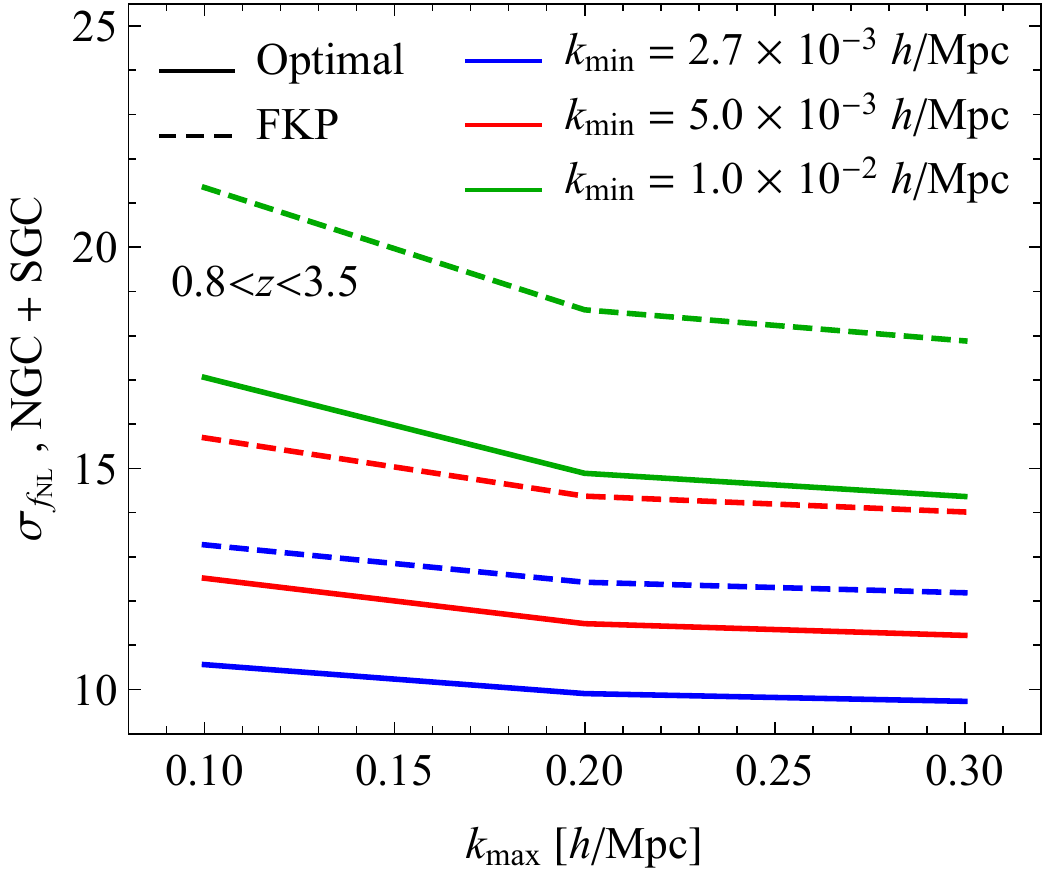}
\hfill
\includegraphics[width=0.49\textwidth]{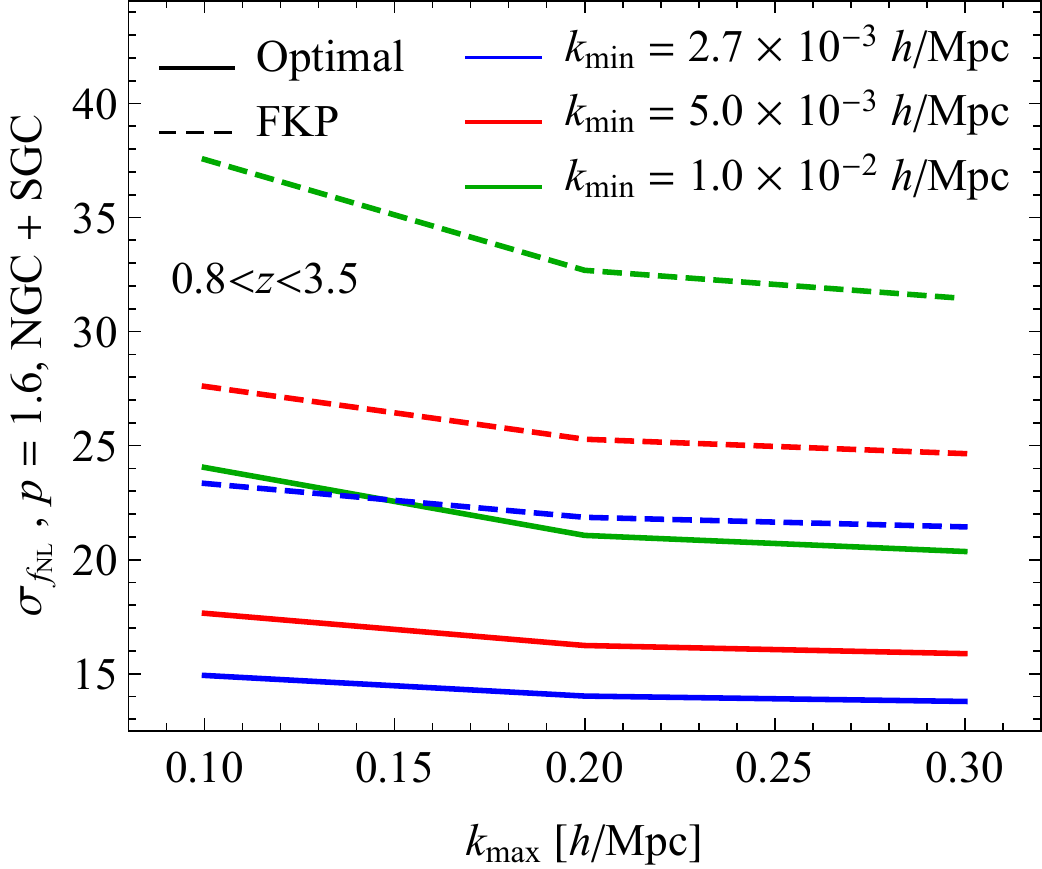}
\caption{Constraints on $\fnl$ using DR14Q in $0.8<z<3.5$, encompassing the full redshift range of eBOSS. Left and right panel correspond to $p=1$ and $p=1.6$ respectively. Note the difference in scale of the $y$-axis between the two panels.}
\label{fig:zmax3p5}
\end{figure}

Figure~\ref{fig:zmax3p5} shows the error on PNG, $\sigma_{\fnl}$, that could be obtained by optimally combining all the eBOSS data between $0.8<z<3.5$. Several choices of $k_{\rm min}$ are displayed, to help understand how the lack of control on systematics on the largest scales affects the final result. 
Similarly to what we find in the nominal DR14 redshift range, the optimal analysis yields 25-30\% improvement for $p=1$, and 40-50\% for $p=1.6$.
We notice that only in the case of the optimal weighting the full survey could in principle achieve $\sigma_{\fnl}\simeq  10$, even at relatively high $k_{\rm min}$. 
The final eBOSS footprint is expected to be roughly 3 times larger than the one of DR14, which could in principle shrink $\sigma_{\fnl}$ by an additional $3^{1/2}$. A survey like eBOSS could therefore achieve an error as low as 
$\sigma_{\fnl}\simeq 5\text{-}8$, depending on the value of $\fnl$ response, if systematics can be kept under control. 

\section{eBOSS DR14 constraints on Primordial Non Gaussianities}\label{sec:constraints}

\begin{figure}
    \centering
    \includegraphics[width=1.\textwidth]{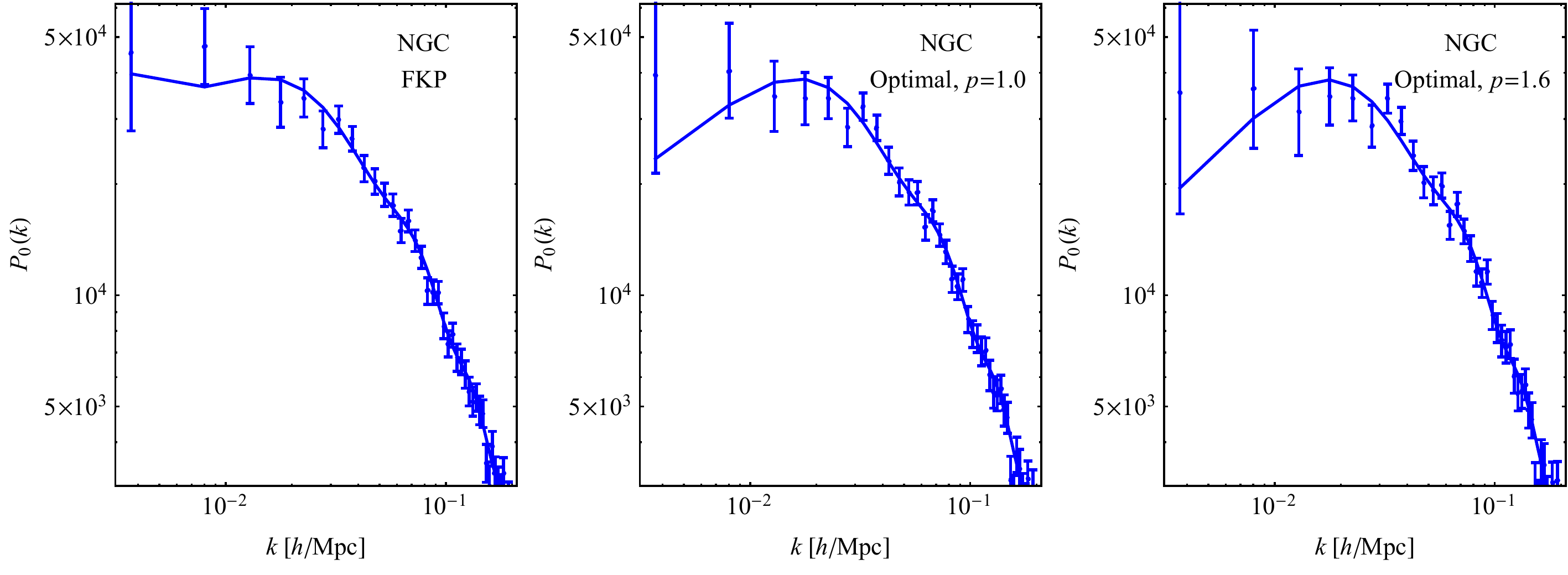}
    \includegraphics[width=\textwidth]{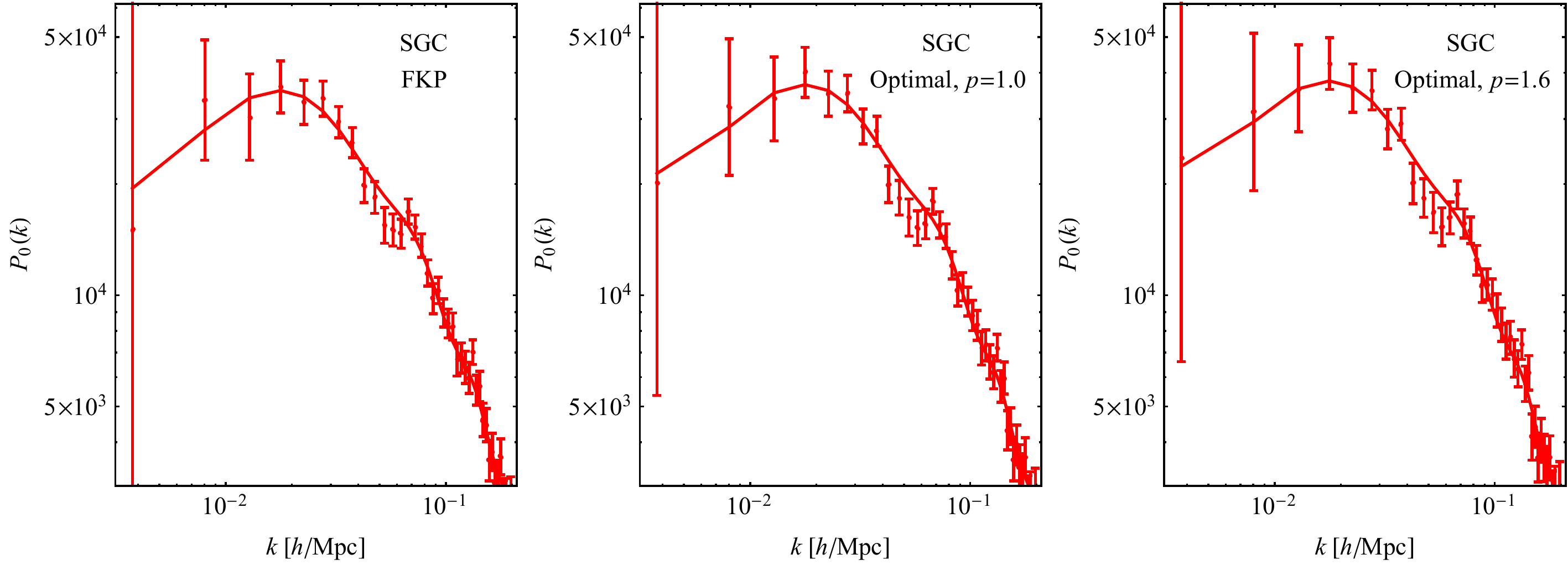}
    \caption{Measurements, points with error bars, and best fit theoretical models, continuous lines, of the monopole power spectrum of the eBOSS DR14 QSOs.
    The top row shows the power spectra in NGC and the lower ones in SGC.}
    \label{fig:data}
\end{figure}

\begin{figure}
    \centering
    \includegraphics[width=0.375\textwidth]{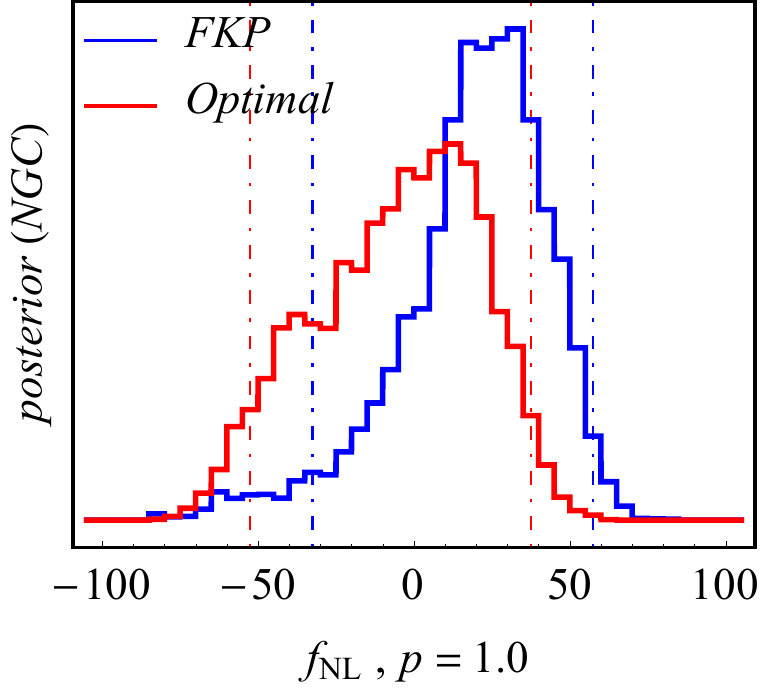}
    \includegraphics[width=0.375\textwidth]{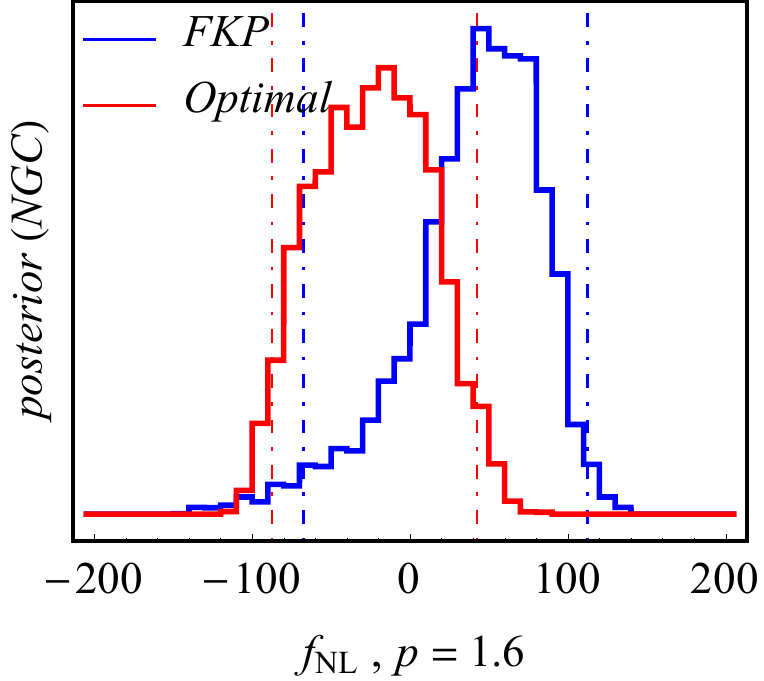}
    \includegraphics[width=0.375\textwidth]{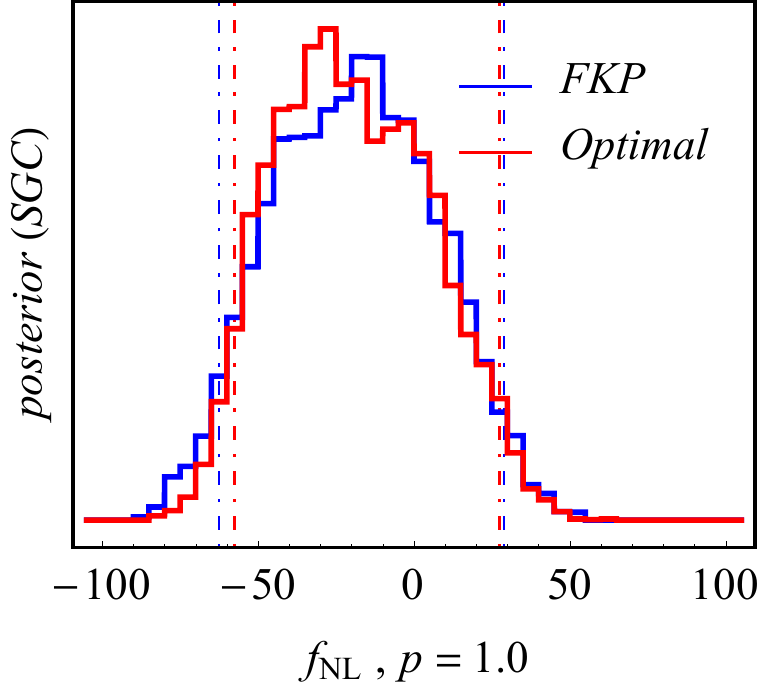}
    \includegraphics[width=0.375\textwidth]{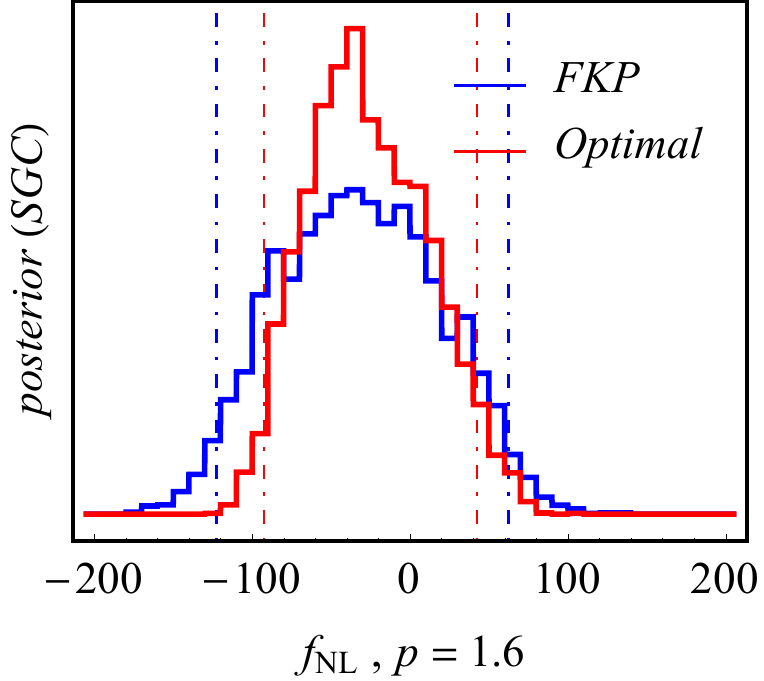}
    \caption{The 1-dimensional posterior of $\fnl$ from separate fits to the NGC (upper panels) and SGC (lower panels). The red histograms show the optimal weighting while the blue ones the FKP weighting. Dot-dashed lines indicate 95\% confidence intervals.}
    \label{fig:NGC_SGC}
\end{figure}

\begin{figure}
    \centering
    \includegraphics[width=0.75\textwidth]{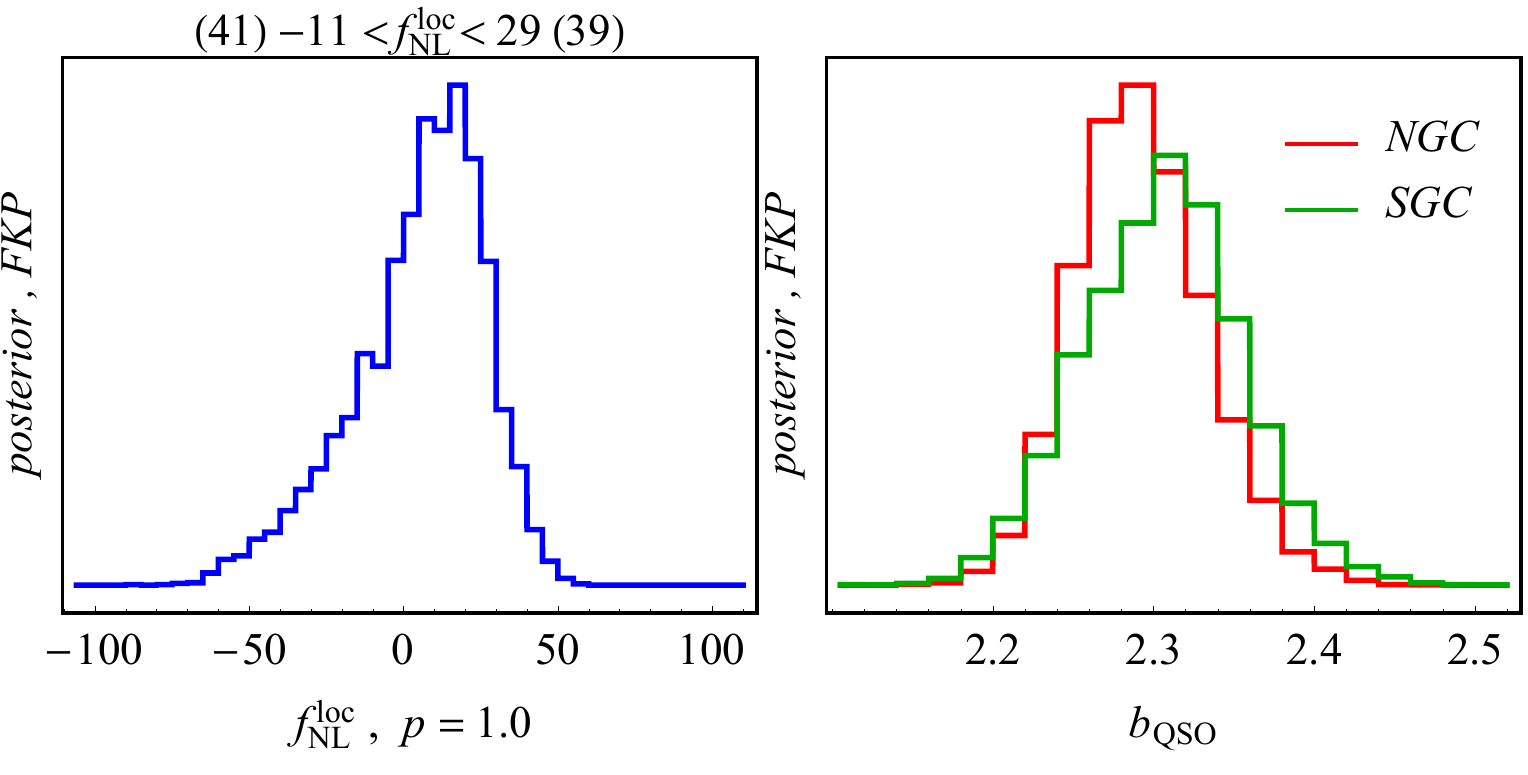}
    \includegraphics[width=0.75\textwidth]{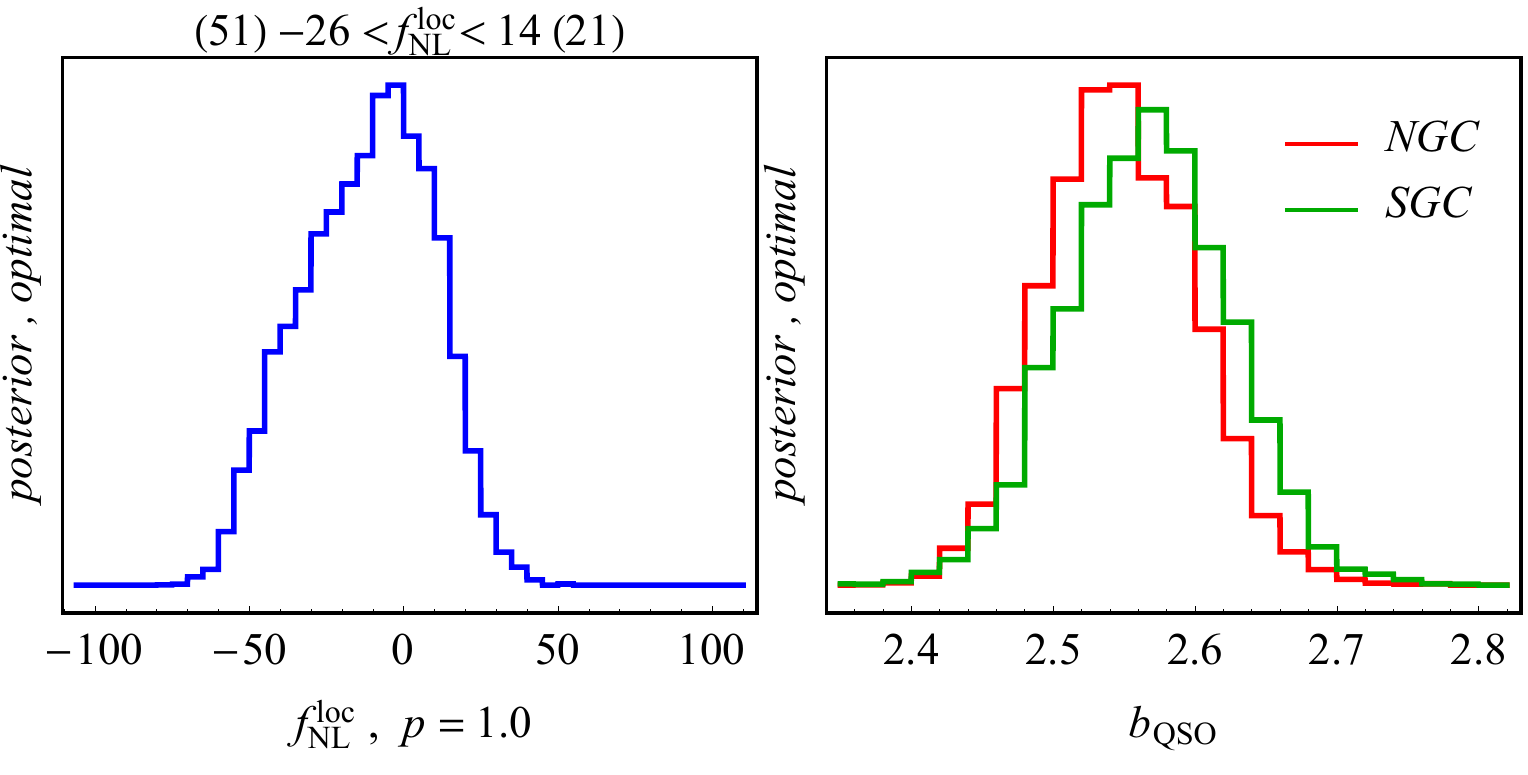}
    \caption{Posterior distribution from jointly fitting the NGC and SGC sky regions, assuming $p=1.0$ for the $\fnl$ response. {\it Left:} 1-dimensional posteriors of $\fnl$ (blue). {\it Right:} The 1-dimensional posteriors for the QSO bias in NGC (red) and SGC (green). The upper panels show the FKP weighting, while the lower ones show the optimal weighting.}
    \label{fig:NGCSGC_p1p0}
\end{figure}

\begin{figure}
    \centering
     \includegraphics[width=0.75\textwidth]{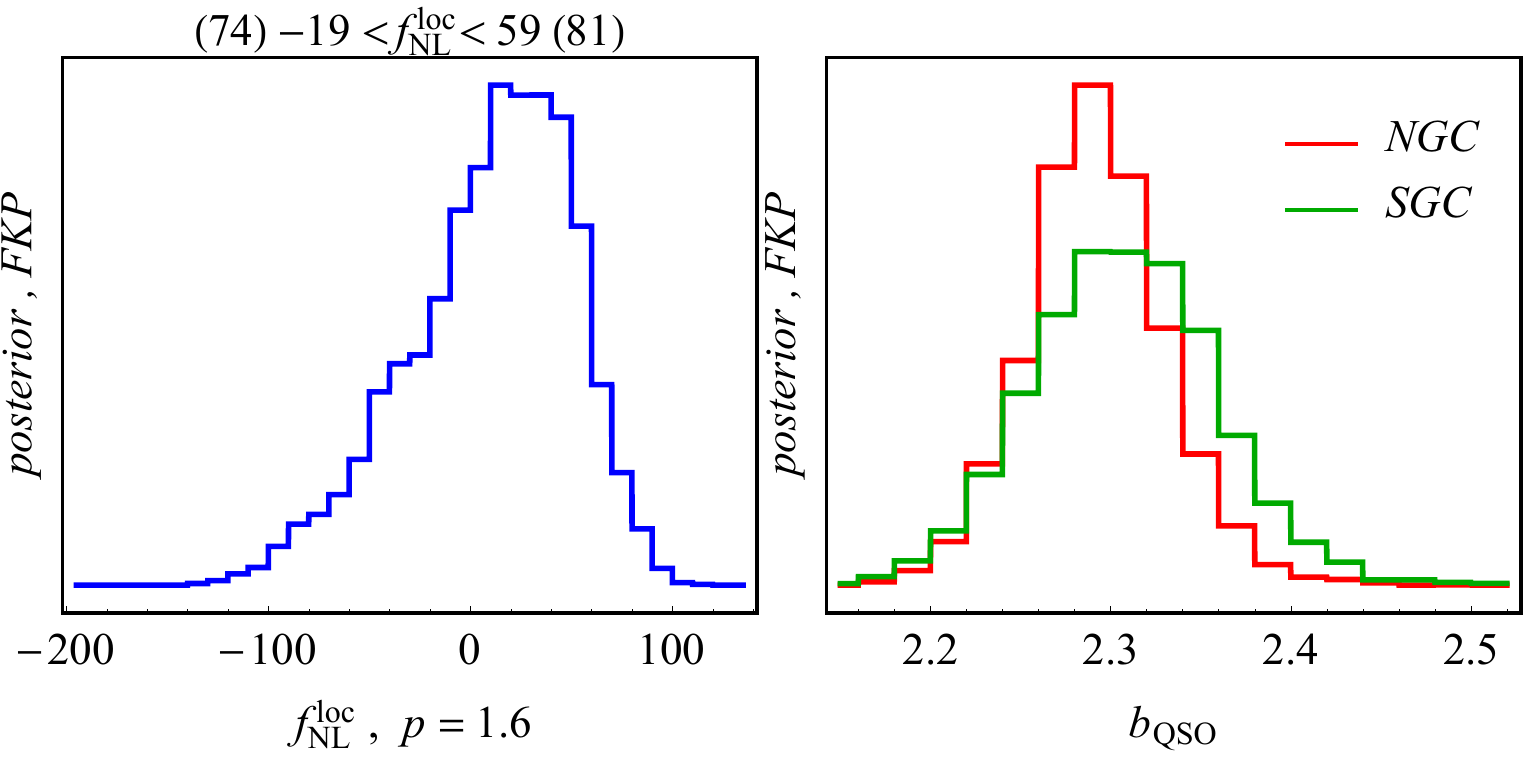}
    \includegraphics[width=0.75\textwidth]{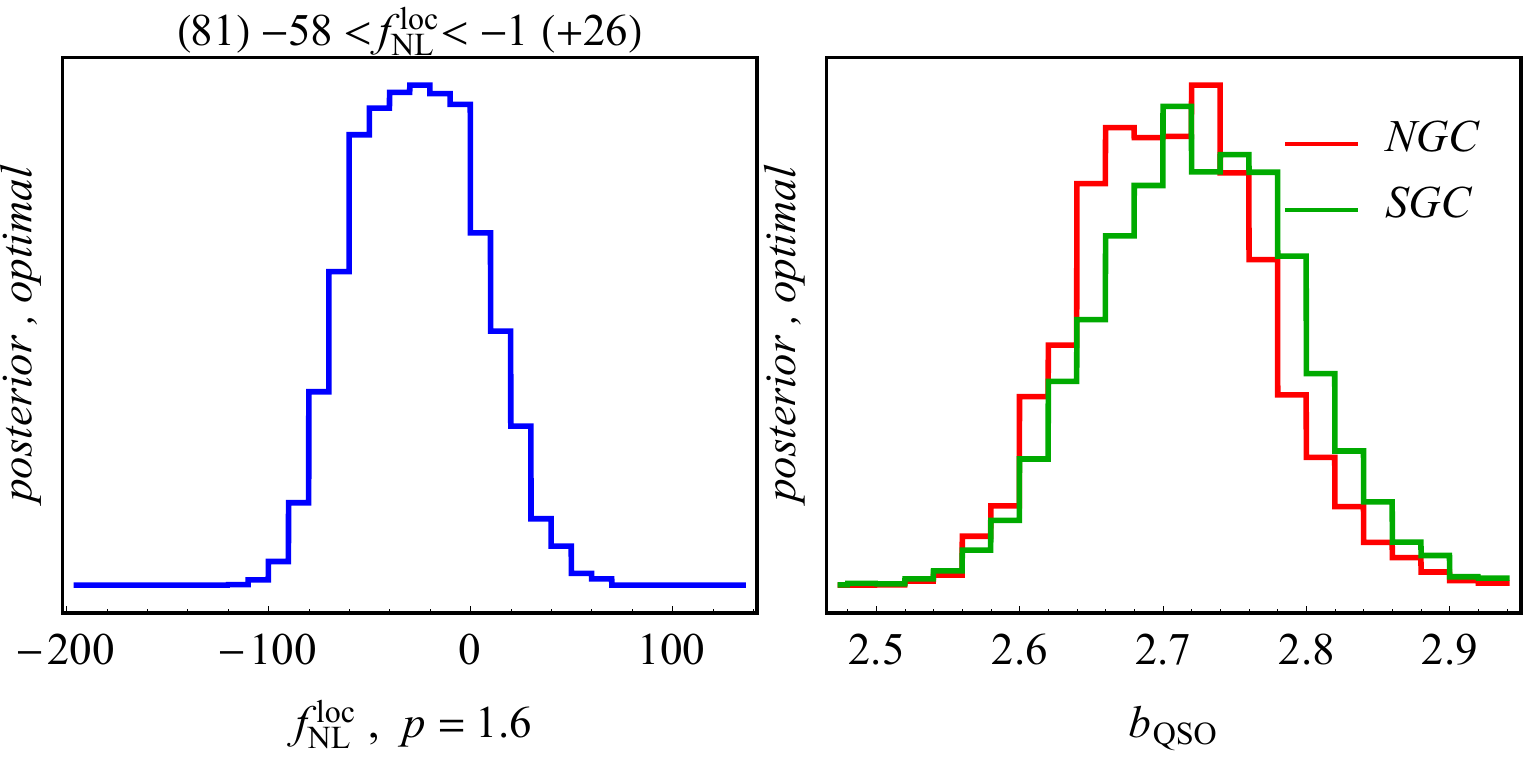}
    \caption{Posterior distribution from jointly fitting the NGC and SGC sky regions, assuming $p=1.6$ for the $\fnl$ response. {\it Left:} 1-dimensional posteriors of $\fnl$ (blue). {\it Right:} The 1-dimensional posteriors for the QSO bias in NGC (red) and SGC (green). The upper panels show the FKP weighting, while the lower ones show the optimal weighting.}
    \label{fig:NGCSGC_0001_p1p6}
\end{figure}

We now move to the analysis of the DR14Q catalog.
Given the different targeting efficiency in the two patches we treat NGC and SGC separately, with two different values of bias, velocity dispersion and shot noise. This implies that the constraints on $\fnl$ will not be as tight as in the idealised Fisher calculation.

The first thing we notice in the measured power spectra is that the redshift-space quadrupole in NGC shows an excess of power on large scales not compatible with a cosmological signal. In this paper we therefore focus on the monopole only, which does not show sign of contamination at low-$k$ that could cause a biased estimate of PNG. It is worth keeping in mind that generically residual foregrounds will affect $\ell=2$ stronger than $\ell=0$, and that systematic effects can only add power to 
the monopole of the power spectrum. So as long as there is no detection, 
an upper limit on $\fnl$ should be reliable.
Since we work at fixed shape of the power spectrum, the quadropole is almost irrelevant for the final constraint on $\fnl$, but could become important if the growth factor, or equivalently $\Omega_m$, is allowed to vary. A brief description of the inconsistency of the quadrupole data can be found in Appendinx \ref{sec:quad}.

In \fig{fig:data} we show the measurements, points with error bars, of the monopole of the power spectrum of eBOSS DR14Q in the two regions of the sky, NGC in the upper plots and SGC in the lower ones.  The three columns correspond to the standard FKP weighting of the data, and the optimal redshift weighting for $p=1.0$ and $p=1.6$.
The best fit models, including $\fnl$ as a free parameter, are displayed as the continuous lines and provide an excellent fit to the data, $\chi^2_{\rm d.o.f.}\simeq1$ in all cases.

\fig{fig:NGC_SGC} shows the one dimensional posterior of $\fnl$ for the different measurements. The first thing worth pointing out is the non-Gaussian posteriors. One expects this because the response to the negative 
$\fnl$ is very different to the positive $\fnl$, as seen in the right panel of Figure~\ref{fig:window-effects}. 
%If this result is not due to residual foregrounds on large scales \citep{Ross:2012}, 
It could be explained by the fact that negative $\fnl$ is a worse fit for FKP weights, and for NGC in particular, see \fig{fig:data}, unless $|\fnl|$ is very large and the bias is much larger than the fiducial value of \cite{Laurent:2017}. 
Non-Gaussian posteriors make the comparison between the FKP and the optimal analysis more difficult, in general there is no unique procedure to compare the two. In all the panels, the  dot-dashed lines correspond to the narrowest region encompassing 95\% of the area under the posterior, the highest posterior density interval. 
%For this choice all values of $\fnl$ outside of the 95\% interval have lower probability than the ones inside. Another popular choice to define bayesian credible intervales is the so called highest posterior density interval which corresponds to the narrowest interval containing 95\% of the area under the posterior. We find that the two definition yields slightly different results at the 95\%, but are basically equivalent at the 99.7\%. 
%The improvement of the optimal analysis compared to the FKP depends on the choice of credible interval. 
The actual numbers at 95\% confidence level for both patches of the sky can be found in Table \ref{tab}.

The optimal analysis, in red, always returns smaller 95\% c.l. intervals, as it can be most easily noticed in the lower right panel.
For $p=1.0$ we find that in NGC the 95\% confidence interval for the optimal analysis is very close to the FKP one (but 10\% smaller at 99.7\%), and it is 5\% smaller in SGC (15\% smaller at 99.7\%). Compared to the Fisher analysis we therefore find smaller gains for the optimal weighting.
For $p=1.6$, the optimal analysis improves considerably over FKP, with the 95\% c.l. now  35-40\% smaller for both NGC and SGC.
Despite NGC being larger than SGC we do not find any appreciable difference between the two in constraining power on PNG, which could be a statistical fluctuation, or an indication of some systematic differences. 
Even though it is not significant enough to bias our results, it should be further investigated in new data releases of eBOSS, which will have better statistics. 

The best possible constraints on $\fnl$ correspond to the joint analysis of NGC and SGC, for which the results are shown in Figure~\ref{fig:NGCSGC_p1p0}. The left panels show the $\fnl$ posterior, while the right
shows QSOs bias parameters for the two areas of the sky.
At 68 (95)\% confidence level we find $(41)\,-11<\fnl<29\,(39)$ for NGC+SGC, $p=1.0$ and FKP weights.  In the optimal case the constraint reads $(51)\,-26<\fnl<14\,(21)$. In the combined NGC+SGC case the FKP analysis is therefore 10\% worse than the optimal one.  
As expected the QSO bias has increased in the optimal case, in accordance with the higher effective redshift of the survey. 
For $p=1.6$ the joint constraints on PNG are shown in \ref{fig:NGCSGC_0001_p1p6}. The improvement of the optimal analysis is more than 35\%. 
It is also important to notice that the optimal weights make the difference between the constraints on $\fnl$ for the two values of $p$ much smaller compared to ones in the FKP case, where $p=1.6$ is almost a factor of 2 worse than $p=1.0$.
Since the true response of any discrete tracer will never be exactly known, our results shows the importance of optimal signal to noise weighting in making this extra source of uncertainties less important.

The numbers given above are among the tightest constraints on PNG using LSS data, and the most stringent one using spectroscopy data of a single tracer. Given the much smaller area of the sky and number of objects compared to \citep{Slosar:2008,Leistedt:2014b}, our analysis strongly indicates the 3D information is crucial to achieve tight constraints on PNG.
At the same time our work makes clear that much more effort should be devoted to studying systematic effects at large angular separation, and remove those modes altogether if a proper foreground cleaning procedure cannot be found \cite{Ross:2012,Leistedt:2014,Hand:2017,Kalus19}.

\begin{figure}
    \centering
    \includegraphics[width=\textwidth]{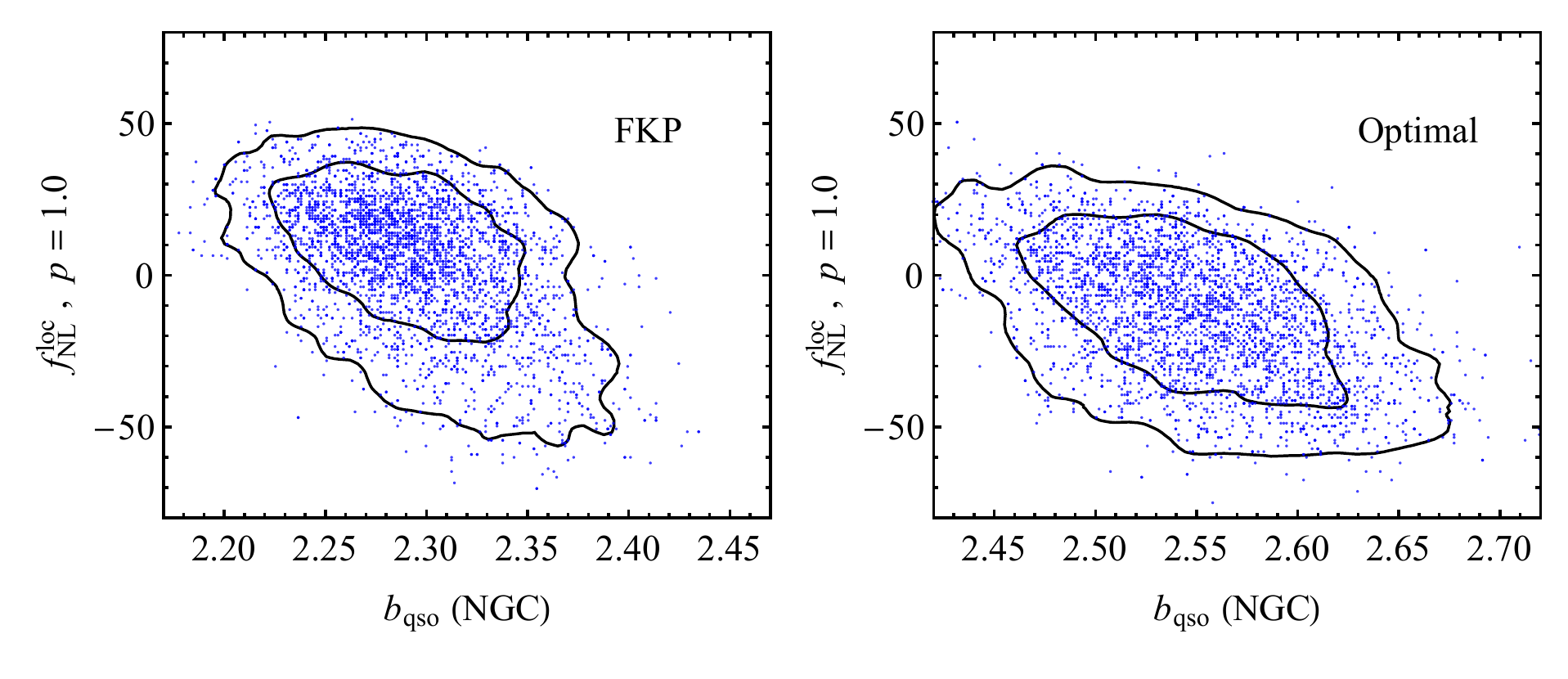}
    \includegraphics[width=\textwidth]{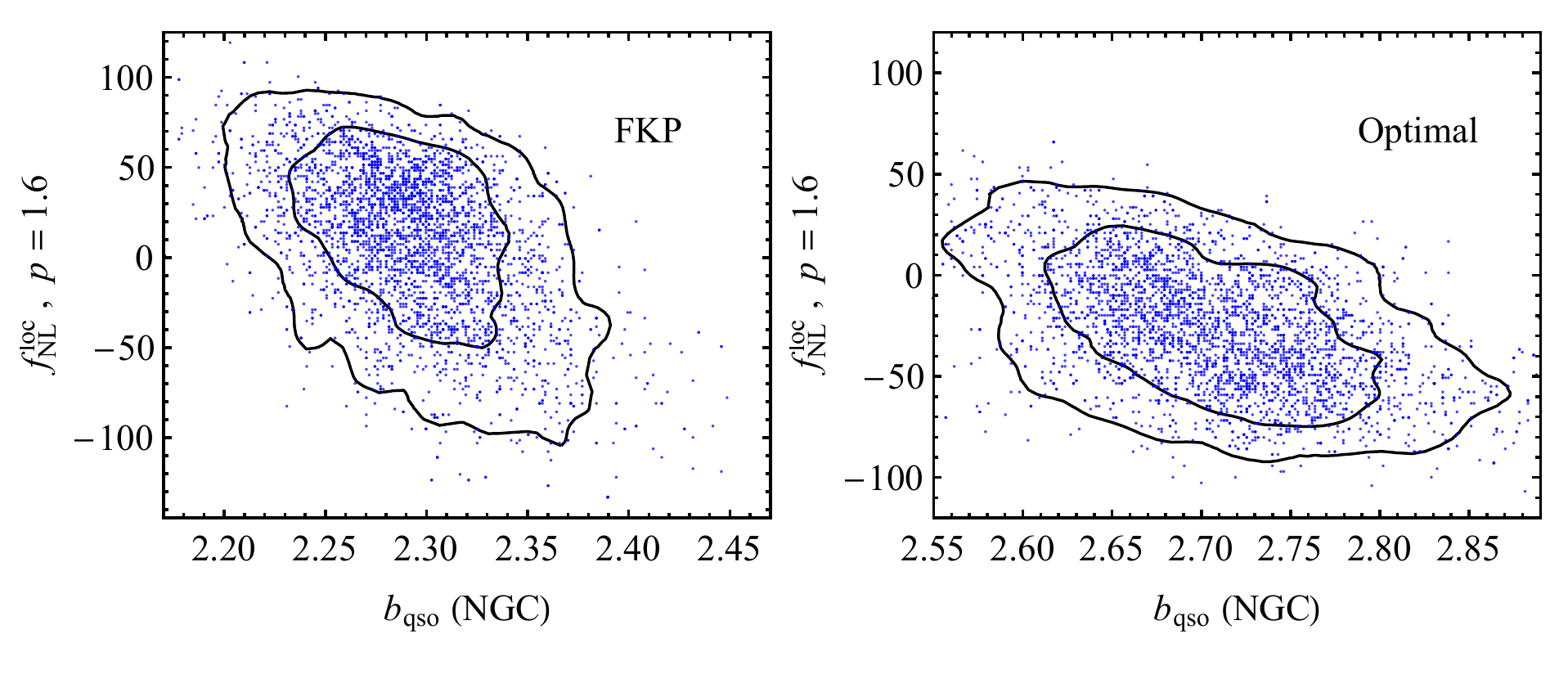}
    \caption{2-dimensional posterior of $\fnl$ and QSOs bias in NGC (similar results hold in SGC). The upper panels assume $p=1.0$, on the right including only FKP weights, on the left with the addition of $\fnl$ optimal weights. Notice how the 2-dimensional contours shrinks in the optimal case, and the $\fnl$ and the QSOs bias become less correlated. The lower set of plots show the same results for $p=1.6$, with exactly the same conclusions.}
    \label{fig:NGCSGC_0001_2d}
\end{figure}

Optimal weights also help in reducing degeneracies between $\fnl$ and other parameters. \fig{fig:NGCSGC_0001_2d} shows the 2-dimensional posterior of $\fnl$ and the QSOs bias in NGC (similar results hold in SGC). The FKP case is shown in the left panels, and for $p=1.0$ and $p=1.6$ in the upper and lower plots, respectively. Comparison with the optimal analysis on the right hand panels shows  that the optimal weights help shrinking the 2-dimensional contours and the correlation between $\fnl$ and the QSOs bias.
A summary of the constraints on $\fnl$ can be found in Table \ref{tab}. We repeated the analysis removing the first $k$-bin, and found an increase of 20-30\% in errors as compared to the analysis including all the bins presented above.

Finally, we would like to comment on the use of mocks to validate the constraints we get from the data.
The EZ mocks have not been tuned to reproduce the $\fnl$ response $b_\phi$ of the eBOSS data, as the latter is unknown.
One would need full physics simulations of the specific eBOSS QSOs sample in order to at least have a theoretical prior on the value of $p$, and consistently compare the constraints on the mocks with the ones obtained in the data.
This implies that, if the PNG response of the sample is poorly determined, mocks can only be used to estimate the covariance matrix of the measurements at the fiducial value of $\fnl=0$. 
We also found that the way the mocks have been generated, with BAO and RSD as their primary goals, does not guarantee that the power spectrum on the largest scales correctly reproduce linear theory. We indeed find a small discrepancy, not significant compared to the expected cosmic variance of the measurements, between the mocks and the theory at the largest scales, that to the best of our knowledge cannot be attribute to modeling systematics.
We were nevertheless able to check that even in the mocks the optimal analysis improves over the standard FKP one, but the actual  mock constraints cannot be compared with the data.

\begin{table}
\begin{center}
\begin{tabular}{ |l|l|l|l| }
\hline
 & & & $\fnl$ Constraint \\
\hline
 & & & \\[0.01in]
\multirow{6}{*}{NGC} &\multirow{3}{*}{$p=1.0$} & FKP & $-34\le\fnl\le61$ \\[0.01in] & & & \\[0.01in]
 
 & & Optimal & $-56\le\fnl\le 38$ \\[0.1in] 
  
   \cline{2-4} & & & \\[0.01in]
  
&\multirow{3}{*}{$p=1.6$} & FKP & $-67\le\fnl\le112$ \\ & & & \\
 
 & & Optimal & $-87\le\fnl\le 42$ \\[0.1in]
 
 \hline \hline & & & \\[0.01in]
 
 \multirow{6}{*}{SGC} &\multirow{4}{*}{$p=1.0$} & FKP & $-64\le\fnl\le31$\\[0.01in] & & & \\[0.01in]
 
 & & Optimal & $-61\le\fnl\le 26$ \\[0.1in]
  
   \cline{2-4}  & & & \\[0.01in]
  
&\multirow{3}{*}{$p=1.6$} & FKP & $-122\le\fnl\le63$ \\ & & & \\
 
 & & Optimal & $-92\le\fnl\le 42$ \\[0.1in]
 \hline \hline & & & \\[0.01in]
 \multirow{6}{*}{NGC+SGC} &\multirow{3}{*}{$p=1.0$} & FKP & $-39\le\fnl\le41$\\[0.01in] & & & \\[0.01in]
 
 & & Optimal & $-51\le\fnl\le 21$ \\[0.1in]
  
   \cline{2-4} & & & \\[0.01in]
  
&\multirow{3}{*}{$p=1.6$} & FKP & $-74\le\fnl\le81$\\ & & & \\
 
 & & Optimal & $-81\le\fnl\le 26$ \\[0.1in] \hline
\end{tabular}

\caption{\label{tab} Summary of the $\fnl$ constraints of this work for NGC and SGC separately, as well as their joint analysis.}
\end{center}
\end{table}

\section{Conclusions}
\label{sec:conclusions}
Primordial non Gaussianities of the local form, parametrized by $\fnl$, leave a unique fingerprint in the clustering of LSS tracers through the presence of scale-dependent bias on large scales.
In this work we presented new constraints on $\fnl$ using the measurements in Fourier space of the clustering of QSOs in DR14 of the eBOSS survey.
In order to access the largest available volume we took all the data in the redshift range $0.8<z<2.2$ without applying any redshift binning. This allowed us to probe modes up to $k = 3.7\times10^{-3} \,\kMpc$.
In such a wide redshift range the evolution of the Gaussian part of the signal is quite significant, and differs from the one of the non-Gaussian piece.

We derive a set of weights that maximizes the information content on PNG in 
the form of a cross correlation between two differently weighted fields, 
a statistically optimal way to exploit the different evolution of the 
two signals. 
Our approach extends the standard FKP weighting, in which all the galaxies are treated the same way from the point of view of their signal content. 
The optimal weights for PNG up-weight higher redshift objects for two main reasons. First they are more highly biased, thus have a higher $\fnl$ response, and second is that the relative size of the dominant Gaussian term to the non-Gaussian piece is smaller at high redshift since the Gaussian term had less time to grow.

In a spectroscopic survey, one in principle has to integrate the expected signal over the redshift distribution of the galaxies for a proper comparison to the data. However, it is quite often a good approximation to introduce an effective redshfit, defined by the $n(z)$ and the desired set of weights $w(z)$. This is the standard assumption in all galaxy survey analyses, and it has been extensively tested for FKP weights.
We checked that the effective redshift approximation is quite accurate even for $\fnl$ optimal weights, and defined two effective redshifts, one for the monopole and one for the quadrupole. 

We then quantified, using a Fisher matrix approach, the expected improvement on $\sigma_{\fnl}$ of the optimal treatment compared to a standard one, finding 15-40\% gain depending on the exact value of the $\fnl$ response.
We also forecasted the possible improvement yielded by including QSOs at $z>2.2$. Our calculation indicates that $\sigma_{\fnl} \simeq 5\text{-}8$ could be obtained by final eBOSS QSOs in the redshift range $0.8<z<3.5$,  if the 
low $k$ systematics can be kept under control. We find no significant 
contamination at low $k$ for the $z<2.2$ QSO sample used in this work.

The exact value of the QSOs response is not known, and can be parametrized by a single number $p$ (higher $p$ means smaller $\fnl$ signal), which the optimal weights depend on. In this work we considered $p=1.0$ and $1.6$, with the 
former value valid if QSOs halo occupation is random, and the latter if QSOs occupy recent merger halos \cite{Slosar:2008}. 
Our current constraints can be summarized as $-51\le\fnl\le21$ at 95\% confidence level for $p=1.0$, and they degrade to $-81\le\fnl\le26$ for $p=1.6$.
It is also worth stressing that the optimal analysis makes the difference between the $p=1.0$ and $p=1.6$ case much smaller than in a standard approach.
The constraints on PNG presented here are some of the tightest ever obtained using tracers of LSS. 

To conclude, in this work we demonstrated the importance of optimal signal weighting in order to extract the maximum information from the data. This required prior analytic knowledge of the signal one is trying to measure, and it reinforces the need for stronger connection between the theory and data
analysis for primordial non Gaussianity. 
We focused on Primordial Non Gaussianities in the power spectrum, but our approach can be straightforwardly extended to any other cosmological parameter and summary statistics. We will return to these interesting questions in future work.

\acknowledgments{
EC thanks Simone Ferraro for useful discussions on Fisher matrices and Andreu Font-Ribera for collaboration in the early stages of this work.
NH is supported by the National Science Foundation Graduate Research Fellowship under
grant number DGE-1106400. US is supported by NASA grant NNX15AL17G, 80NSSC18K1274 and NSF 1814370, NSF 1839217.
Funding for the Sloan Digital Sky Survey IV has been provided by the Alfred P. Sloan Foundation, the U.S. Department of Energy Office of Science, and the Participating Institutions. SDSS-IV acknowledges
support and resources from the Center for High-Performance Computing at
the University of Utah. The SDSS web site is www.sdss.org.

SDSS-IV is managed by the Astrophysical Research Consortium for the 
Participating Institutions of the SDSS Collaboration including the 
Brazilian Participation Group, the Carnegie Institution for Science, 
Carnegie Mellon University, the Chilean Participation Group, the French Participation Group, Harvard-Smithsonian Center for Astrophysics, 
Instituto de Astrof\'isica de Canarias, The Johns Hopkins University, 
Kavli Institute for the Physics and Mathematics of the Universe (IPMU) / 
University of Tokyo, the Korean Participation Group, Lawrence Berkeley National Laboratory, 
Leibniz Institut f\"ur Astrophysik Potsdam (AIP),  
Max-Planck-Institut f\"ur Astronomie (MPIA Heidelberg), 
Max-Planck-Institut f\"ur Astrophysik (MPA Garching), 
Max-Planck-Institut f\"ur Extraterrestrische Physik (MPE), 
National Astronomical Observatories of China, New Mexico State University, 
New York University, University of Notre Dame, 
Observat\'ario Nacional / MCTI, The Ohio State University, 
Pennsylvania State University, Shanghai Astronomical Observatory, 
United Kingdom Participation Group,
Universidad Nacional Aut\'onoma de M\'exico, University of Arizona, 
University of Colorado Boulder, University of Oxford, University of Portsmouth, 
University of Utah, University of Virginia, University of Washington, University of Wisconsin, 
Vanderbilt University, and Yale University.
P.D. C. and G.R. acknowledge support from the National Research Foundation of Korea (NRF) through Grant No. 2017077508 funded by the Korean Ministry of Education, Science and Technology (MoEST), and from the faculty research fund of Sejong University in 2018.
}

\appendix
\section{The power spectrum quadropole}
\label{sec:quad}
\begin{figure}
    \centering
    \includegraphics[width=0.5\textwidth]{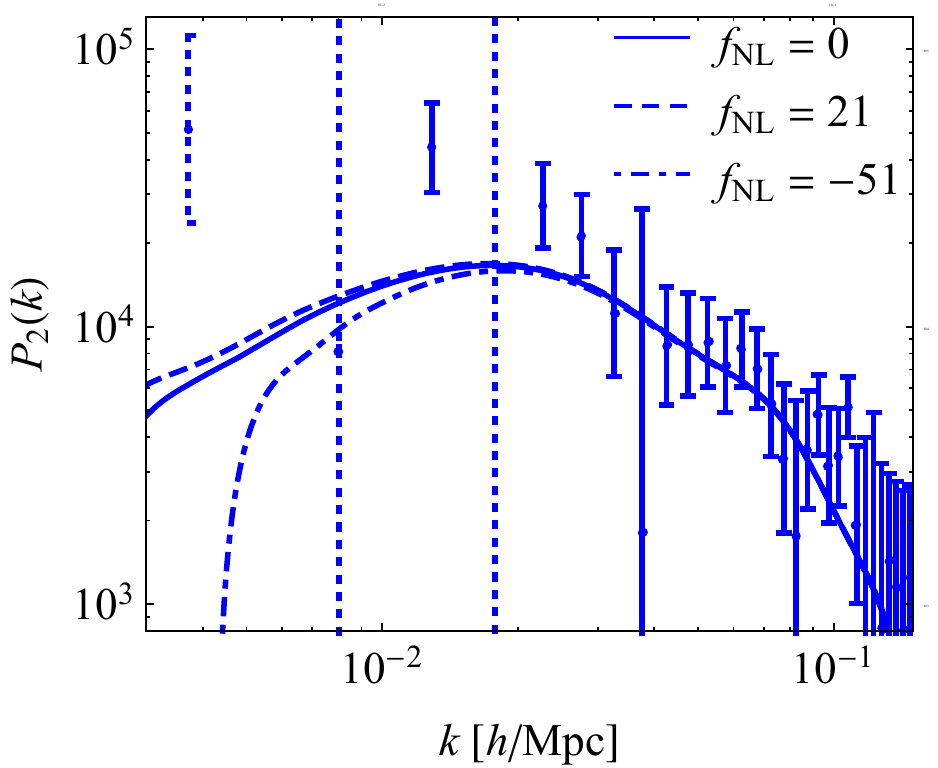}
    \caption{The measurement of the power specturm quadrupole in NGC, point with errorbars (dotted for negative values), in comparison with the model prediction for $\fnl=0,-51,21$. }
    \label{fig:P2_NGC}
\end{figure}

In the main text we included only the monopole of the power spectrum in the data analysis. This was motivated by the anomalous excess power we observe in the quadropole of the data. In \fig{fig:P2_NGC} we show the power spectrum quadropole in NGC, point with errorbars (dotted for negative values), compared to the theoretical prediction for $\fnl=0$ and $\fnl=-51,21$, which correspond to the $\pm95\%$ values of $\fnl$ in the analysis on the monopole data.
Clearly even for such high value of PNG the quadropole data are inconsistent with the theoretical model, and should therefore be neglected due to their covariance with the monopole measurements.

\bibliographystyle{JHEP}
\bibliography{}
\end{document}